\documentclass[aps,prl,twocolumn,groupedaddress,amsmath,epsfig,amssymb]{revtex4}
\usepackage{makecell}
\usepackage{bbm}
\usepackage{mathrsfs}
\usepackage{subfigure}
\usepackage{lipsum} 
\usepackage{amsfonts}
\usepackage{color}
\usepackage{threeparttable}
\usepackage{multirow}
\usepackage{graphicx}
\usepackage{float}
\usepackage[normalem]{ulem}
\def\newblock{\hskip .11em plus .33em minus .07em}

\newcommand{\rv}{{\mathbf r}}

\newcommand{\uv}{{\mathbf u}}

\newcommand{\vv}{{\boldsymbol v}}
\newcommand{\kv}{{\mathbf k}}

\newcommand{\xh}{{\hat{x}}}
\newcommand{\yh}{{\hat{y}}}

\newcommand{\kh}{{\hat{k}}}

\newcommand{\be}{\begin{equation}}
\newcommand{\ee}{\end{equation}}
\newcommand{\ba}{\begin{eqnarray}}
\newcommand{\ea}{\end{eqnarray}}

\begin{document}
	\title{Alignment Destabilizes Crystal Orders in Active Systems}
	\author{Chen Huang$^{1}$}
	\author{Leiming Chen$^{2}$}
	\author{Xiangjun Xing$^{1,3,4}$}
	\email{xxing@sjtu.edu.cn}
	\address{$^1$ Wilczek Quantum Center, School of Physics and Astronomy, Shanghai Jiao Tong University, Shanghai 200240 China\\
$^2$ School of Materials Science and Physics,
China University of Mining and Technology, Xuzhou, Jiangsu, 221116  China\\
	$^3$ T.D. Lee Institute, Shanghai Jiao Tong University, Shanghai 200240 China \\
	$^4$ Shanghai Research Center for Quantum Sciences, Shanghai 201315 China}
	
	
	
\begin{abstract}
We combine numerical and analytical methods to study two dimensional active crystals formed by permanently linked swimmers and with two distinct alignment interactions.  The system admits a stationary phase with quasi long range translational order, as well as a moving phase with quasi-long range velocity order.  The translational order in the moving phase is significantly influenced by alignment interaction.  For Vicsek-like alignment, the translational order is short-ranged, whereas the bond-orientational order is quasi-long ranged, implying a moving hexatic phase.  For elasticity-based alignment, the translational order is quasi-long ranged parallel to the motion and short-ranged in perpendicular direction, whereas the bond orientational order is long-ranged.  We also generalize these results to higher dimensions. 
\end{abstract}

\maketitle

{\bf Introduction} \quad One of the most interesting and fundamental issues about active systems~\cite{toner2005hydrodynamics,ramaswamy2010mechanics,vicsek2012collective,marchetti2013hydrodynamics,ginelli2016physics,chate2020dry} is the stability of orders.  According to the Mermin-Wagner theorem~\cite{mermin1966absence}, two-dimensional (2D) equilibrium systems with continuous symmetry and short range interaction cannot exhibit not long range order (LRO).  However, LRO was discovered in 2D polar active fluid both in simulation~\cite{Vicsek-1995} and in hydrodynamic theory~\cite{toner1995long,toner2012reanalysis,tu1998sound,toner1998flocks}.  This LRO is accompanied by super-diffusion and giant number fluctuations~\cite{tu1998sound, gregoire2004onset, ramaswamy2010mechanics,chate2020dry}, neither of which is seen in equilibrium systems with short range interactions.  Many variants of Vicsek models with different particle polarity, alignments and exclusion~\cite{toner1998flocks,chate2006simple, ramaswamy2003active, mishra2006active,ginelli2010large,weber2013long,bhattacherjee2019re,weber2014defect,martin2018collective,sese2018velocity,gregoire2004onset,mishra2010dynamic,shankar2018low,tasaki2020hohenberg,levis2017synchronization} have been studied, a variety of novel phenomena have been discovered.    

 
\vspace{-1mm}

Dense active systems with repulsive interactions may also exhibit translational orders.   Solid phases as well as fluid-solid phase separations have been repeatedly observed in active colloidal systems both experimentally~\cite{theurkauff2012dynamic,buttinoni2013dynamical,palacci2013living} and numerically~\cite{gregoire2003moving,fily2012athermal,bialke2013microscopic,redner2013structure,bialke2012crystallization}.  In most of these works, there is no alignment interaction, and no visible collective motion.  More recently, Weber {\em et. al.}~\cite{weber2014defect}\footnote{Note however in this work the dynamic equations contain no noise term.  } simulated a model active crystal with Vicsek-type alignment, and discovered a stationary phase with quasi-long range (QLR) translational order, as well as a phase of moving crystal domains separated by grain boundaries.   Very recently, Maitra {\it et.al.}~\cite{maitra2019oriented} studied an active generalization of nematic elastomer~\cite{warner2007liquid} with spontaneous breaking of rotational symmetry~\cite{xing2003thermal,xing2008nonlinear}, and found QLR translational orders in 2D.  Their elastic energy contains a hidden rotational symmetry (and its resulting Goldstone modes) involving both shear deformation and orientational order, which are difficult to realize experimentally. 


\begin{figure}[t]
	\centering
	\includegraphics[width=3.5in]{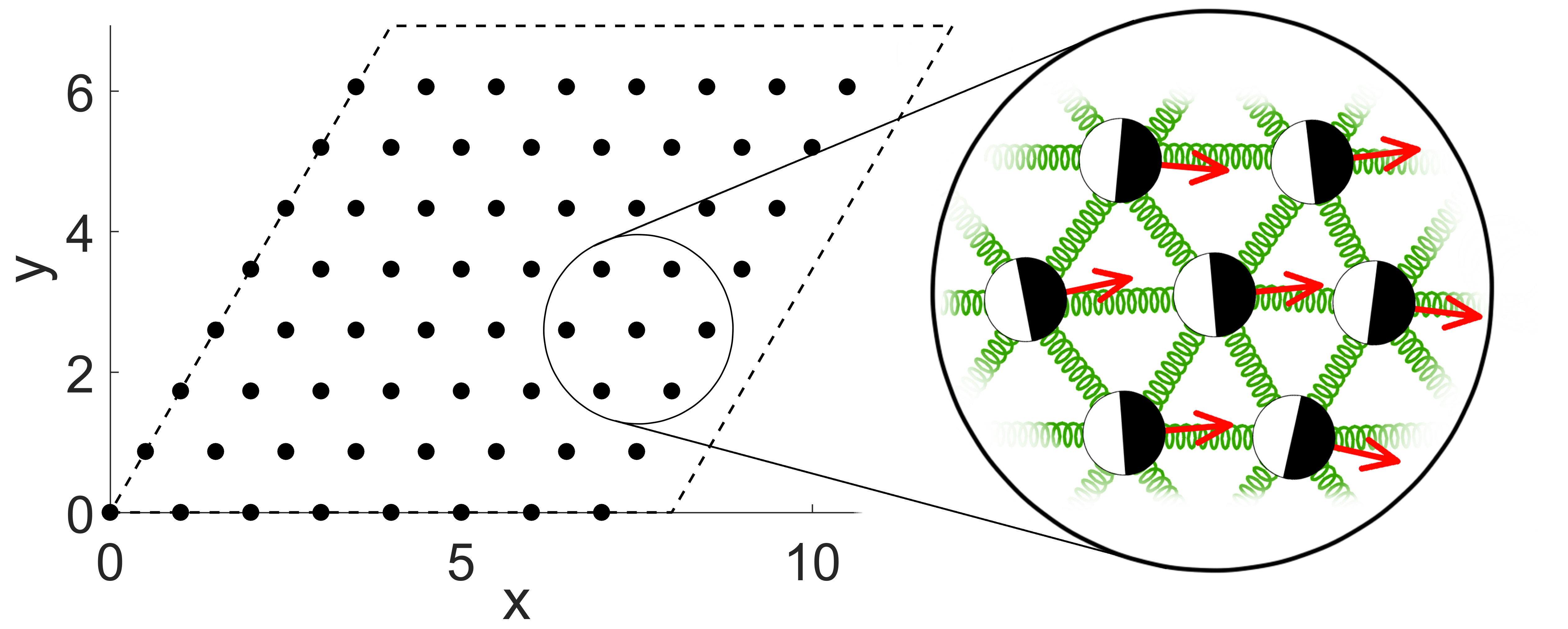}
	\caption{Our simulation model.  Swimmers are connected by springs and driven by active forces, shown as red arrows.  } 
	\label{fig:illustration}
\vspace{-5mm}
\end{figure}

Regardless of many previous studies, it is not clear whether there exists a moving phase with certain translational order in active systems alignment interactions, if the soft-mode in Ref.~\cite{maitra2019oriented} does not come into play.  To address this interesting question, here we combine analytic and numerical approaches to study a model system of active crystal consisting of 2D triangular array of swimmers linked permanently by springs.   We introduce alignment interaction between neighboring swimmers that is either Vicsek-like (AD-I) or elasticity-based~\cite{ferrante2013elasticity,ferrante2013collective} (AD-II).  In the strong noise/weak alignment regime, we find a stationary phase with QLR translational order, which was also seen in Ref.~\cite{weber2014defect}.  In the weak noise/strong alignment regime we find a moving phase with QLR velocity order, and with the nature of translational order depending on the alignment.  For Vicsek-like alignment (AD-I), the moving phase exhibits only short-range (SR) translational order and QLR bond orientational order, and hence should be identified with {\em moving hexatic phase}.  For elasticity-based alignment (AD-II), the translational order is QLR along the moving direction, and SR in the perpendicular direction, whereas the bond-orientational order is LR.  We generalize the model to higher dimensions, and show in active systems velocity alignment tends to destabilize crystal orders.

{\bf Simulation model }\quad As schematized in Fig.~\ref{fig:illustration}, our simulation model consists of a triangular array of swimmers connected permanently by harmonic springs.  Each swimmer moves under the influences of elastic force, friction and noise, as well as active force.  The position ${\mathbf{r}}_k(t)$ of $k$-th swimmer in the lab frame then obeys the following over-damped Langevin equation:
\ba
\gamma \, \dot{\mathbf{r}}_k(t)
=  b\, \hat{n}(\theta_k) + \mathbf{F}_k(t)
+ \, \gamma \sqrt{2 D}  \, \hat{\xi}_k(t),
\label{LD_discrete}
\ea
where $\gamma$ is the friction coefficient, $ \hat n (\theta_k) =(\cos\theta_k,\sin\theta_k)$ is  the director of active force.  The magnitude of active force $b$ is assumed to be fixed in our model.  The elastic force is
\be
\mathbf{F}_k = \sum_{j \, n.n. k}
\kappa( |\mathbf{r}_{k}-\mathbf{r}_{j} |-a_0)
\frac{ \mathbf{r}_{k}-\mathbf{r}_{j}}{ |\mathbf{r}_{k}-\mathbf{r}_{j}|}, 
\label{elastic-force-def}
\ee 
where the summation is over six nearest neighbors of swimmer $k$,  whilst $\kappa$ and $a_0$ are respectively the elastic constant and natural length of the springs.  The last term of Eq.~(\ref{LD_discrete}) is the random force, with $\hat{\xi}_k(t)$ the unit Gaussian white noise.  
 

We consider two distinct alignment dynamics for the active forces.  The first (AD-I) is Vicsek-like~\cite{vicsek1995novel}, with each swimmer trying to align its director of active force with its neighbors~\footnote{Note that in the original Vicsek model~\cite{vicsek1995novel}, swimmers control their velocities instead of active forces.  In the over-damped regime, this difference is inessential.  }, subject to an internal noise:  
\ba
\dot{\theta}_k(t)&=&  d (\hat{n}({\theta}_k) \times 
\langle \hat {n} \rangle_k ) 
\cdot \hat{z} + \sqrt{2D_{\theta}} \, \eta_k(t), 
\quad \text{(AD-I)}
\nonumber
\ea
where $\hat{z}$ is the unit normal to the plane, $\langle \hat {n} \rangle_k  \equiv \sum_{j \, {\rm n.n.} k} \hat{n} (\theta_j)/6$ the average director of all six nearest neighbors, and  $ \eta_k(t)$ is a unit Gaussian white noise.

The second alignment dynamics (AD-II) is elasticity-based~\cite{ferrante2013elasticity,ferrante2013collective,huepe2015scale}, with each swimmer aligning its active force with the local elastic force, so as to reduce the local elastic energy: 
\ba
\dot{\theta}_k(t) &=& c (\hat{n}({\theta}_k)  \times \mathbf{F}_k) 
\cdot \hat{z} + \sqrt{2D_{\theta}} \, \eta_k(t).  
\quad \text{(AD-II)}
\nonumber
\ea

We  use a rhombic cell with periodic boundary condition (c.f. Fig.~\ref{fig:illustration}), and  numerically integrate the dynamic equations, Eqs.~(\ref{LD_discrete}) and (AD-I) or (AD-II), using the first-order Euler-Maruyama scheme~\cite{kloeden2013numerical}, with the time step $\Delta\tilde{t}=10^{-3}$.  Simulation details  as well as the definitions of all dimensionless parameters, are given in Supplementary Information (SI) Sec. I.  

\begin{figure}[t!]
	\centering
	\includegraphics[width=3.3in]{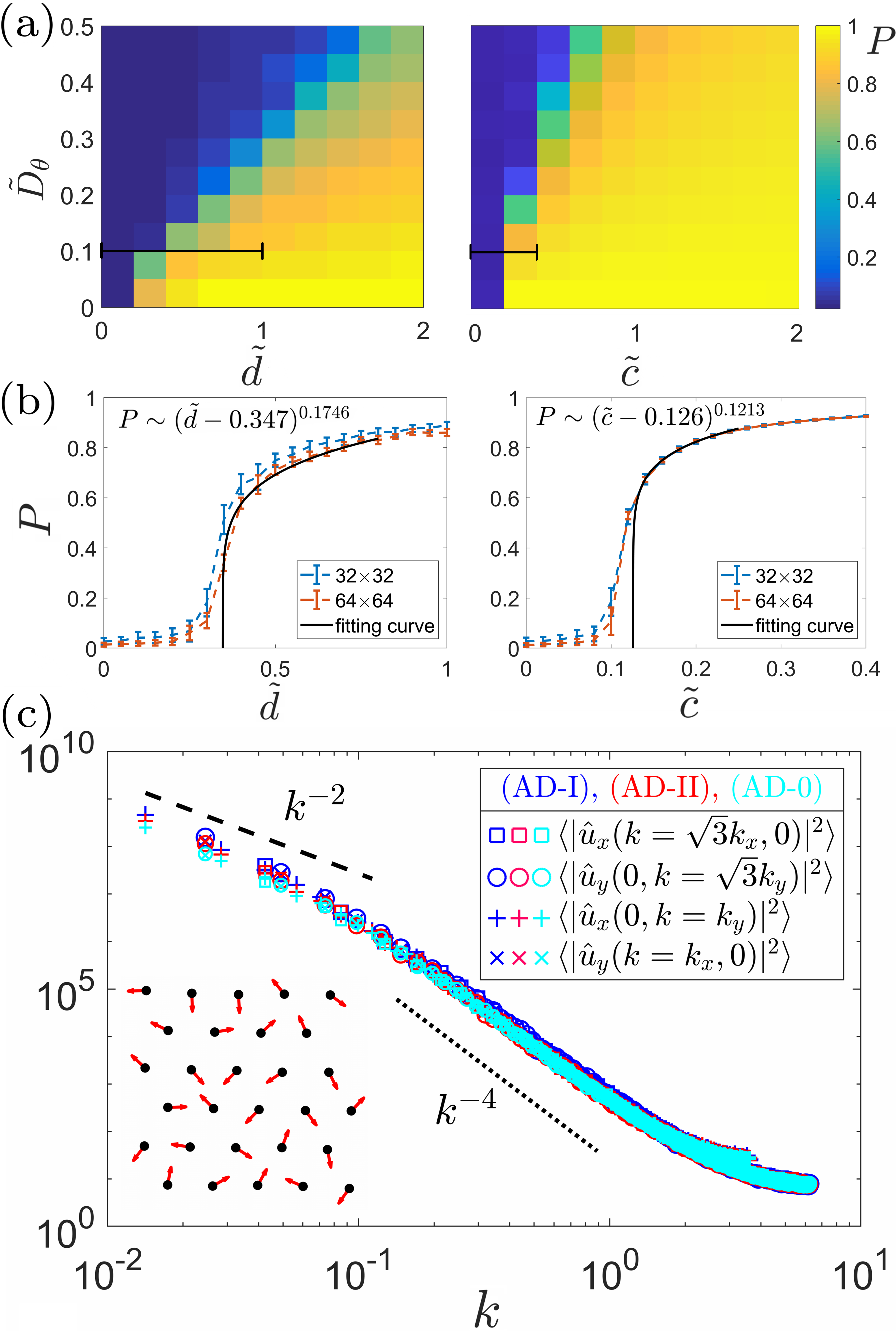}
	\caption{(a) Phase diagram of AD-I (left) and AD-II (right).  Vertical axis: dimensionless noise strength $\tilde{D}_{\theta}$; horizontal axis: dimensionless alignment $\tilde{d}$ (AD-I) or $\tilde{c}$ (AD-II). The color represents velocity order parameter $P$, defined in the man text.  {(b) Velocity order parameter as a function of alignment, suggesting second-order phase transitions. }  Left (AD-I), right (AD-II).  (c) Log-log plot of the phonon correlation functions $\langle |\hat{u}_x|^2\rangle$ and $\langle |\hat{u}_y|^2\rangle$ of the stationary phase along the $\kh_x$ and $\kh_y$ axis, with parameters: $\tilde{b}=2$, $\tilde{D}=0.01$, $\tilde{D}_{\theta}=0.3$, $(\tilde{\kappa}, \tilde{d}) = (100, 0.1)$ in AD-I, $(\tilde{\kappa}, \tilde{c}) = (20, 0.1)$ in AD-II, and $(\tilde{\kappa}, \tilde{d}, \tilde{c}) = (100, 0, 0)$ in AD-0. Bottom left inset: active forces of the stationary phase.
}
	\label{fig::phaseDiagram}
	\vspace{-5mm}
\end{figure}

We first determine the phase diagram by computing the velocity order parameter, defined as the steady state time average of magnitude of system-averaged active force ${P} =\langle | \frac{1}{N} \sum_k \hat n(\theta_k) | \rangle_t$, which is proportional to the velocity of collective motion.  As shown in Fig.~\ref{fig::phaseDiagram}(a), for weak alignment/strong active noise (dark blue in  upper left) there is a stationary phase where ${P}$ is approximately zero~\footnote{It is never strictly zero because the system is finite and there are always instantaneous fluctuations at each time.}, whereas for strong alignment/weak active noise (bright yellow in lower right) there is a collectively moving phase where $P$ is finite.  As shown in Fig.~\ref{fig::phaseDiagram}(b), fitting of $P$ as a function of alignment strength suggests that these two phases are separated by a line of second order phase transitions.





To study the translational order in the stationary phase, we carry out a larger simulation with system size $256 \times 256$. The total number of time steps is $2\times10^6$ and simulation samples are collected every $2000$ steps in the steady state.  We Fourier transform the phonon field, and compute averages of their norm squared: $\langle |\hat{u}_x(\kv)|^2\rangle$, $\langle |\hat{u}_y(\kv)|^2\rangle$, which are often called as {\em phonon correlation functions} (in momentum space).   The technical details of numerical computation are presented in SI Sec. II.  In Fig.~\ref{fig::phaseDiagram}(c) we plot $\langle |\hat{u}_x(\kv)|^2\rangle$ and $\langle |\hat{u}_y(\kv)|^2\rangle$ along $\kh_x$ and $\kh_y$ directions in $\kv$ space, and  both for  AD-I and for AD-II  in log-log scale.   For comparison we also plot the corresponding result for the model without any alignment (AD-0), which corresponds to $\tilde{d}=\tilde{c}=0$ in Eqs.~(AD-I) and (AD-II).  It is remarkable that  all curves collapse onto each other, and exhibit $k^{-4}$ scaling in the intermediate length scale ($0.1< k < 1$).  This signify anomalously large structure fluctuations in the length scales $6 a_0  < \ell < 60 a_0$ ($a_0$ the lattice constant)  which are caused by the fluctuations of active forces.  These results strongly suggest that alignment does not play a role in the stationary phase.  For smaller $k$ the phonon correlation functions crossover to $k^{-2}$ scaling, indicating a QLR translation order, in consistent with the result obtained in Ref.~\cite{weber2014defect}.   



\begin{figure}[t!]
	\centering
	\includegraphics[width=3.4in,height=4.8in]{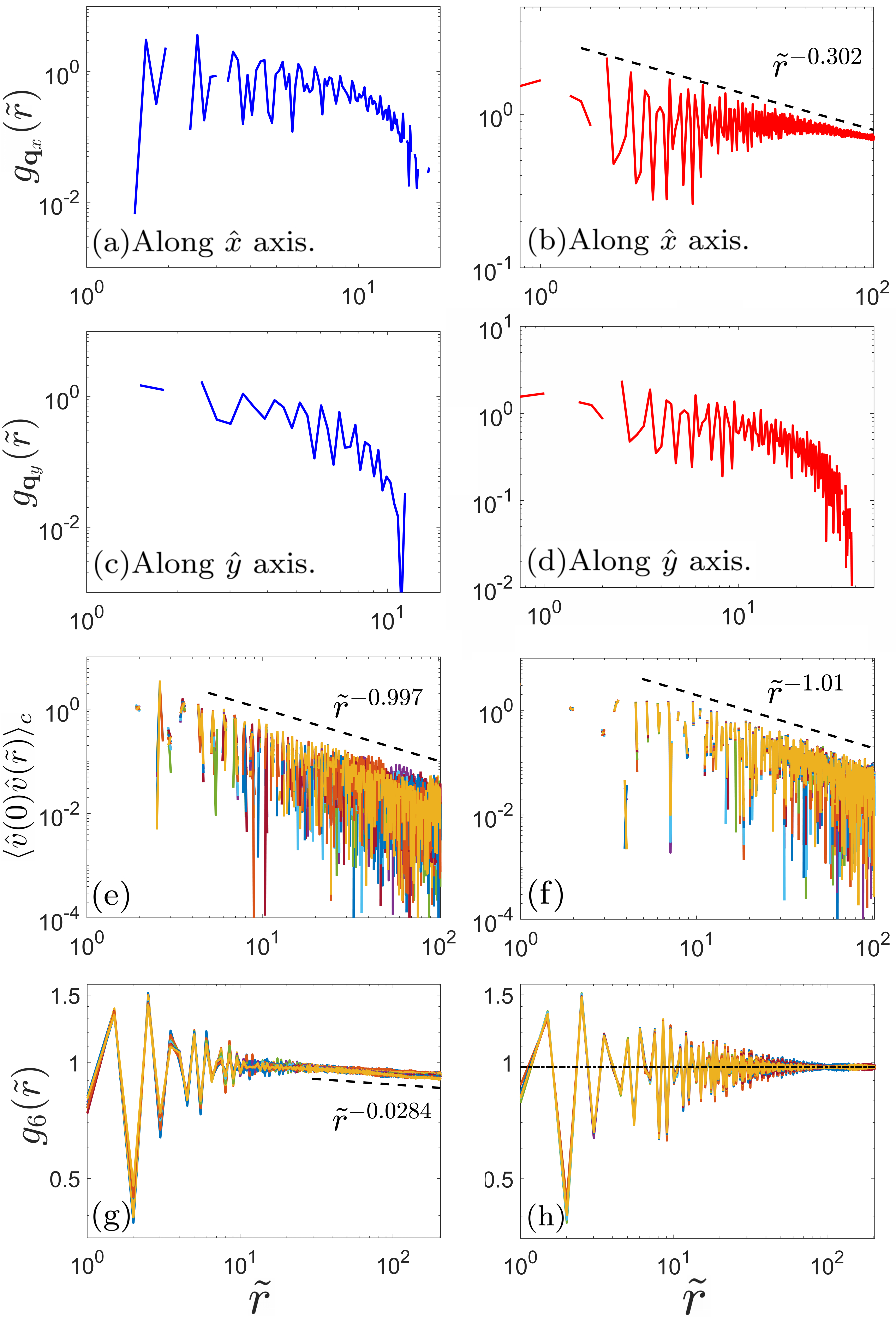}
	
\caption{(a)-(d) Translational correlation functions; (e) \&(f) velocity correlation functions; (g)\&(h) bond-orientational correlation functions.  Left column: AD-I, right volume: AD-II.  The parameters are $\tilde{D}_{\theta}=0.05$, $(\tilde{\kappa},\tilde{d})=(100,10)$ for AD-I and $(\tilde{\kappa},\tilde{c})=(20,1)$ for AD-II. 
We start from an initial state with perfect triangular lattice and $\theta_k=0$. Data are collected after the system reaches a steady state with $P\approx 0.9915$ in AD-I and $P\approx 0.9830$ in AD-II.}
	\label{fig::AD_II_transCorr}
	\vspace{-5mm}
\end{figure}

We are however most interested in the  collective moving phase.  For this purpose, we study the translational correlation function $g_{\mathbf{q}}(\tilde{r})$, velocity correlation function $\langle \hat v(0) \hat v(\tilde{r}) \rangle_c$, and the bond-orientational correlation function $g_6(\tilde{r})$, all of which are defined defined in SI Sec. III.   In particular, $g_{\mathbf{q}}(\tilde{r})$, defined in Eq.~(12) of SI Sec. III,  characterizes the correlation of translational order with Bragg vector $\mathbf q$ at two regions separated by a dimensionless distance $\tilde{r}\equiv r/a_0$~\footnote{Here the orientation of real space displacement is averaged over.  The directionality of $g_{\mathbf{q}}(\tilde{r})$ is carried by the Bragg vector $\mathbf q$.  For details, see SI Sec. III. }.  As demonstrated in Ref.~\cite{li2019accurate}, because of sample-to-sample fluctuations of crystal structure, the Bragg peak $\mathbf{q}$ used for $g_{\mathbf{q}}(\tilde{r})$ must be identified carefully for each sample state.  In Fig.~\ref{fig::AD_II_transCorr}(a) \& (c) we plot $g_{\mathbf{q}}(\tilde{r})$ for AD-I along $\xh$ and $\yh$ axes for a typical state of the moving phase, which clearly  decays faster than power law.  Hence there is only  short-range (SR) translational order in the moving phase of AD-I.  The corresponding results for AD-II are shown in Fig.~\ref{fig::AD_II_transCorr}(b) \& (d), where it is seen that $g_{\mathbf{q}}(\tilde{r})$ decays algebraically along the $\hat{x}$ axis, and decays faster along the $\hat{y}$ axis.   In Fig.~\ref{fig::AD_II_transCorr}(e) \&(f) we show that velocity correlation functions decay in power-law, which indicate QLRO of velocity field for both AD-I and AD-II.  As displayed in Fig.~4(c) \& (d) of  SI Sec. III, we also find QLRO for the director correlation of active forces.  Furthermore, it appears that the exponent of all these power-law scalings are very close to unity.  In Fig.~\ref{fig::AD_II_transCorr}(g) \& (h), we show that the bond-orientational correlation function $g_6(\tilde{r})$ decays in power law for AD-I and converges to a finite limit for AD-II.  Hence  the bond orientational order is quasi-long ranged for AD-I and long-ranged for AD-II.  

In summary, the moving phase of AD-I exhibits QLRO both in velocity and in bond-orientation, but only has SR translational order.  Hence it should be categorized as a {\em moving hexatic phase}.  By contrast, the moving phase of AD-II exhibits QLRO in velocity and LRO n bond-orientation, yet the translation order is quasi-long ranged along the moving direction and short ranged in the other direction.  This resembles the active smectic phase~\cite{adhyapak2013live,chen2013universality}, even though there is no visible layer structures in our system.  These numerical results indicate that there is no enhancement of velocity order due to translation order.  On the other hand, alignment interactions tend to destabilize translational order in active crystals, and that the destabilizing effect is stronger for Vicsek-like alignment (AD-I) than for elasticity-based alignment (AD-II).

\vspace{1mm}
{\bf Analytic treatment} \quad Our lattice model of active crystal can be coarse-grained to yield a continuous theory in terms of phonon fields $\uv(\rv, t)$ and director field $\theta(\rv, t)$. Detailed analyses is given by SI Sec. IV.  Further linearizing, the elastic force Eq.~(\ref{elastic-force-def}) becomes 
\ba
\left\{\begin{array}{ll}
F_{x}&=(\lambda+2 \mu) \partial_x^{2} u_{x} + \mu \partial_y^{2} u_{x} + (\lambda+\mu) {\partial_x \partial_y u_{y}}, 
\vspace{2mm}
\\
F_{y}&=(\lambda+2 \mu) {\partial^{2}_y u_{y}} + \mu {\partial^{2}_x u_{y}}
+ (\lambda+\mu) {\partial_x \partial_y u_{x}},
\end{array}\label{eq:explicitElasticForce}
\right.
\ea
where $\lambda, \mu$ are two Lam\'e coefficients characterizing the solid elasticity.  The dynamics of AD-I is described  by:
\ba
\left\{\begin{array}{ll}
	\gamma \, \dot{u}_x&=F_x+\sqrt{2\gamma T} \, \xi_x\\
	\gamma \, \dot{u}_y&=b \, \theta+F_y+\gamma\sqrt{2D}\, \xi_y\\
	\dot{\theta} &=d \,\Delta\theta+\sqrt{2D_\theta}\, \eta
	\label{eq:peturb_model1-0}
\end{array}.
\right.
\ea
where $\xi_x(\mathbf{r},t), \xi_y(\mathbf{r},t), \eta(\mathbf{r},t)$ are all unit Gaussian white noises.   We Fourier transform Eqs.~(\ref{eq:peturb_model1-0}), and calculate the steady state correlation function for ${\mathbf{u}}(\kv, t)$ and $\theta(\kv, t)$:
\begin{subequations}
\ba
\langle |\hat{u}_x(\kv)|^2\rangle &=&
 \frac{W_1(\alpha)}{k^6} 
+ \frac{W_2(\alpha)}{k^2}, 
\label{correlation-full-ADI-ux}
\ea
\ba
\langle |\hat{u}_y(\kv)|^2\rangle &=& 
\frac{W_3(\alpha)}{k^6} 
+ \frac{W_4(\alpha)}{k^2},
\label{correlation-full-ADI-uy}\\
\langle |\hat{\theta}(\kv)|^2\rangle &=& 
\frac{D_\theta}{d k^2},
\label{correlation-full-ADI-theta}
\ea
\label{correlation-full-ADI}
\end{subequations} 
where $\alpha = \tan^{-1} (k_y/k_x) $ is the polar angle of $\kv$, and the functions $W_i(\alpha), i = 1,2,3,4 $ are defined in Eqs.~(28)  of SI Sec. IV.  While $W_2(\alpha), W_3(\alpha), W_4(\alpha)$ are all positive, 
\ba
\begin{aligned}
	W_1(\alpha) &= \frac{b^2 D_\theta (d\gamma + 4\lambda) \sin^2 2\alpha }
	{ 12 d \lambda (d^2 \gamma^2 + 4 d \gamma \lambda + 3\lambda^2)}.
\end{aligned}
\ea
 vanishes along $\kh_x$ and $\kh_y$ axes, where $\alpha=0, \pi/2, \pi, 3\pi/2$ respectively. Hence along these axes, $\langle |\hat{u}_x(\kv)|^2\rangle$ scales as $k^{-2}$ instead of $k^{-6}$ along other directions. 

The real space fluctuations of phonon fields and director field can be obtained by integrating  Eqs.~(\ref{correlation-full-ADI}) over $\kv$.  Intregrating Eq.~(\ref{correlation-full-ADI-theta}) we see that $\langle \theta(\rv, t)^2 \rangle$, diverges logarithmically with system size, and hence the active force exhibits QLRO.  Since the velocity is massively coupled to the active force, it should also exhibit QLRO, as demonstrated by our numerical simulation. On the other hand, Integrating Eqs.~(\ref{correlation-full-ADI-ux}), (\ref{correlation-full-ADI-uy}) we see that $\langle u_x(\rv, t)^2 \rangle$ and  $\langle u_y(\rv, t)^2 \rangle$ diverge in power law with system size, indicating SR translation orders.  These results again agree with our numerical observations shown in Fig.~\ref{fig::AD_II_transCorr}.



The dynamics of AD-II in the linearized continuous theory can be similarly obtained:
\ba
\left\{\begin{array}{ll}
	\gamma \, \dot{u}_x&= F_x+\gamma\sqrt{2D}\, \xi_x\\
	\gamma \, \dot{u}_y&= b \, \theta+F_y+\gamma\sqrt{2D} \, \xi_y\\
	\dot{\theta}&= c\, F_y + \sqrt{2D_\theta}\, \eta
\end{array}.
\right.
	\label{eq:peturb_model2-0}
\ea
The steady state correlation functions, calculated in detail in SI Sec. IV, turn out to be more complicated.  Here we only display the leading order terms for small $k$: 
\begin{subequations}
\ba
\langle |\hat{u}_x(\kv)|^2\rangle &=&
 \frac{W_5(\alpha)}{k^2} + \mathcal{O}(k^0),
 \label{correlation-full-ADII-ux}
\\ 
\langle |\hat{u}_y(\kv)|^2\rangle &=&
 \frac{W_6(\alpha)}{k^4} + \mathcal{O}({k^{-2}}),
 \label{correlation-full-ADII-uy}
\\
\langle |\hat{\theta}(\kv)|^2\rangle &=& 
\frac{W_7(\alpha)}{k^2} + \mathcal{O}(k^0),
 \label{correlation-full-ADII-theta}
\ea
\label{correlation-full-ADII}
\end{subequations}
where $W_5(\alpha), W_6(\alpha), W_7(\alpha)$ are all positive definite.  Since  Eq.~(\ref{correlation-full-ADII-theta}) scales the same as Eq.~(\ref{correlation-full-ADI-theta}) we see that the velocity order is again quasi-long ranged.  Integrating Eqs.~(\ref{correlation-full-ADII-ux}), (\ref{correlation-full-ADII-uy}) we see that that translational order for AD-II is QLR along $x$ axis and SR along $y$ axis,  again consistent with the numerical results displayed in Fig.~\ref{fig::AD_II_transCorr}.  In SI Sec. V we compare the contour plots of analytical and numerical correlation functions in $\kv$ plane for both active dynamics, see quantitative agreements.  


To achieve better understanding of our model of active solids, we can generalize the analytic theory to arbitrary $d$ dimensions.  The elasticity of a $d$ dimensional isotropic solid can by obtained by adapting Eq.~(\ref{eq:explicitElasticForce}), if we replace $u_y$ and $\partial_y$ by $\uv_{\perp}$ and $\nabla_{\perp}$ where $\perp$ denotes the subspace perpendicular to the moving direction $\hat x$, replace $\theta $ by $\delta \hat n_{\perp}$, the fluctuation of active force director in the perpendicular subspace. Equations (\ref{eq:peturb_model1-0}) and (\ref{eq:peturb_model2-0}) can be similarly generalized to $d$ dimensions.  In SI Sec. VI, an explicit derivation is given for the case $d = 3$.  The correlation functions for the phonon fields and for the director fluctuation $\delta \hat n_{\perp}$ can be similarly computed, and the results are still given by Eqs.~(\ref{correlation-full-ADI}) and (\ref{correlation-full-ADII}), as long as $u_y$ are replaced by $\uv_{\perp}$ and $\theta$ by $\delta \hat n_{\perp}$.   Hence we conclude that for Vicsek-like alignment (AD-I), the translational order of active crystal has critical dimension $d^t_c = 6$, whilst the velocity order has critical dimension $d^v_c = 2$.  By contrast, for elasticity-based alignment (AD-II), the translational order along the moving direction has critical dimension $d_c^{t\, \parallel} = 2$, and that perpendicular to the moving direction has critical dimension $d_c^{t, \perp} = 4$, whilst the velocity order has critical dimension $d_c^{\rm v} = 2$.  The nature of bond-orientational order however can not be easily determined from our analytic theory.  Nonetheless, given our 2D results, we deduce the bond orientational order is long-range above two dimensions both for AD-I and AD-II.  The nature of various orders for 2d and 3d cases are summarized in Table \ref{tabel}. 

\begin{table}[]
	\begin{tabular}{|c|c|c|c|c|c|c|c|}
		\hline
		\multicolumn{2}{|c|}{\multirow{2}{*}{\begin{tabular}[c]{@{}c@{}} \end{tabular}}} & \multicolumn{2}{c|}{Stationary Phase} & \multicolumn{4}{c|}{ Moving Phase} \\ \cline{3-8} 
		\multicolumn{2}{|c|}{} & \quad\quad $\uv$ \quad\quad & $\psi_6$  & $\uv_\parallel$ & \quad $\uv_\perp$ \quad & $\vv$ & $\psi_6$ \\ \hline
		\multirow{2}{*}{2D} 
		& AD-I & QLRO & LRO & SRO & SRO & QLRO & QLRO \\ \cline{2-8} 
		& AD-II & QLRO & LRO & QLRO & SRO & QLRO & LRO\\ \hline
		\multirow{2}{*}{3D} 
		& AD-I & LRO & LRO & SRO & SRO & LRO & LRO \\ \cline{2-8} 
		& AD-II & LRO & LRO & LRO & SRO & LRO & LRO\\ \hline
	\end{tabular}
	\caption{Stability of translational order, bond-orientational orders, and velocity order in active crystals in 2D and 3D.  In the stationary phase, bond-orientational order is always long-ranged whereas velocity is always short-ranged.}
\vspace{-4mm}
		\label{tabel}
\end{table}

In this work, we have provided a complete characterization of various orders in active solids. To test our theoretical results using experiments, sufficiently large system sizes must be prepared and studied.  Both the permanent springs and the alignment interaction may, for instance, be realized by remote sensing between nano-robotics~\cite{turgut2008self,ccelikkanat2010steering,ferrante2010flocking}. It is also conceivable to manipulate self-propelled colloidal particles~\cite{van2019interrupted,lavergne2019group} or link swimmers using polymers and to induce alignment interaction using hydrodynamic effects or magnetic interactions.  It would be more interesting to study whether the phases we discovered are related to those solid-like moving structures observed in Vicsek-type models with short-range interactions~\cite{martin2018collective,sese2018velocity}.   These questions will be explored in a future work.

X.X. acknowledge support from NSFC via grant \#11674217, as well as additional support from a Shanghai Talent Program.  This research is also supported by Shanghai Municipal Science and Technology Major Project (Grant No.2019SHZDZX01).


\begin{thebibliography}{10}
	
	\bibitem{toner2005hydrodynamics}
	John Toner, Yuhai Tu, and Sriram Ramaswamy.
	\newblock Hydrodynamics and phases of flocks.
	\newblock {\em Annals of Physics}, 318(1):170--244, 2005.
	
	\bibitem{ramaswamy2010mechanics}
	Sriram Ramaswamy.
	\newblock The mechanics and statistics of active matter.
	\newblock {\em Annual Review of Condensed Matter Physics}, 1(1):323--345, 2010.
	
	\bibitem{vicsek2012collective}
	Tam{\'a}s Vicsek and Anna Zafeiris.
	\newblock Collective motion.
	\newblock {\em Physics reports}, 517(3-4):71--140, 2012.
	
	\bibitem{marchetti2013hydrodynamics}
	M~Cristina Marchetti, Jean-Fran{\c{c}}ois Joanny, Sriram Ramaswamy,
	Tanniemola~B Liverpool, Jacques Prost, Madan Rao, and R~Aditi Simha.
	\newblock Hydrodynamics of soft active matter.
	\newblock {\em Reviews of Modern Physics}, 85(3):1143, 2013.
	
	\bibitem{ginelli2016physics}
	Francesco Ginelli.
	\newblock The physics of the vicsek model.
	\newblock {\em The European Physical Journal Special Topics},
	225(11):2099--2117, 2016.
	
	\bibitem{chate2020dry}
	Hugues Chat{\'e}.
	\newblock Dry aligning dilute active matter.
	\newblock {\em Annual Review of Condensed Matter Physics}, 11:189--212, 2020.
	
	\bibitem{mermin1966absence}
	N~David Mermin and Herbert Wagner.
	\newblock Absence of ferromagnetism or antiferromagnetism in one-or
	two-dimensional isotropic heisenberg models.
	\newblock {\em Physical Review Letters}, 17(22):1133, 1966.
	
	\bibitem{Vicsek-1995}
	T. Vicsek, A. Czirók, E. Ben-jacob, I. Cohen, and O. Shochet,
	\newblock Novel Type of Phase Transition in a System of Self-Driven Particles, 
	\newblock {\em Phys. Rev. Lett.}, 75, 1226 (1995).
	
	\bibitem{toner1995long}
	John Toner and Yuhai Tu.
	\newblock Long-range order in a two-dimensional dynamical xy model: how birds
	fly together.
	\newblock {\em Physical review letters}, 75(23):4326, 1995.
	
	\bibitem{toner2012reanalysis}
	John Toner.
	\newblock Reanalysis of the hydrodynamic theory of fluid, polar-ordered flocks.
	\newblock {\em Physical Review E}, 86(3):031918, 2012.
	
	\bibitem{tu1998sound}
	Yuhai Tu, John Toner, and Markus Ulm.
	\newblock Sound waves and the absence of galilean invariance in flocks.
	\newblock {\em Physical review letters}, 80(21):4819, 1998.

	\bibitem{gregoire2004onset}
	Guillaume Gr{\'e}goire and Hugues Chat{\'e}.
	\newblock Onset of collective and cohesive motion.
	\newblock {\em Physical review letters}, 92(2):025702, 2004.
	
	\bibitem{toner1998flocks}
	John Toner and Yuhai Tu.
	\newblock Flocks, herds, and schools: A quantitative theory of flocking.
	\newblock {\em Physical review E}, 58(4):4828, 1998.
	
	\bibitem{chate2006simple}
	Hugues Chat{\'e}, Francesco Ginelli, and Ra{\'u}l Montagne.
	\newblock Simple model for active nematics: Quasi-long-range order and giant
	fluctuations.
	\newblock {\em Physical review letters}, 96(18):180602, 2006.
	
	\bibitem{ramaswamy2003active}
	Sriram Ramaswamy, R~Aditi Simha, and John Toner.
	\newblock Active nematics on a substrate: Giant number fluctuations and
	long-time tails.
	\newblock {\em EPL (Europhysics Letters)}, 62(2):196, 2003.
	
	\bibitem{mishra2006active}
	Shradha Mishra and Sriram Ramaswamy.
	\newblock Active nematics are intrinsically phase separated.
	\newblock {\em Physical review letters}, 97(9):090602, 2006.
	
	\bibitem{ginelli2010large}
	Francesco Ginelli, Fernando Peruani, Markus B{\"a}r, and Hugues Chat{\'e}.
	\newblock Large-scale collective properties of self-propelled rods.
	\newblock {\em Physical review letters}, 104(18):184502, 2010.
	
	\bibitem{weber2013long}
	Christoph~A Weber, Timo Hanke, J~Deseigne, S~L{\'e}onard, Olivier Dauchot,
	Erwin Frey, and Hugues Chat{\'e}.
	\newblock Long-range ordering of vibrated polar disks.
	\newblock {\em Physical review letters}, 110(20):208001, 2013.
	
	\bibitem{bhattacherjee2019re}
	Biplab Bhattacherjee and Debasish Chaudhuri.
	\newblock Re-entrant phase separation in nematically aligning active polar
	particles.
	\newblock {\em Soft matter}, 15(42):8483--8495, 2019.
	
	\bibitem{weber2014defect}
	Christoph~A Weber, Christopher Bock, and Erwin Frey.
	\newblock Defect-mediated phase transitions in active soft matter.
	\newblock {\em Physical review letters}, 112(16):168301, 2014.
	
	\bibitem{martin2018collective}
	Aitor Mart{\'\i}n-G{\'o}mez, Demian Levis, Albert D{\'\i}az-Guilera, and
	Ignacio Pagonabarraga.
	\newblock Collective motion of active brownian particles with polar alignment.
	\newblock {\em Soft matter}, 14(14):2610--2618, 2018.
	
	\bibitem{sese2018velocity}
	Elena Sese-Sansa, Ignacio Pagonabarraga, and Demian Levis.
	\newblock Velocity alignment promotes motility-induced phase separation.
	\newblock {\em EPL (Europhysics Letters)}, 124(3):30004, 2018.
	
	\bibitem{mishra2010dynamic}
	Shradha Mishra, R~Aditi Simha, and Sriram Ramaswamy.
	\newblock A dynamic renormalization group study of active nematics.
	\newblock {\em Journal of Statistical Mechanics: Theory and Experiment},
	2010(02):P02003, 2010.
	
	\bibitem{shankar2018low}
	Suraj Shankar, Sriram Ramaswamy, and M~Cristina Marchetti.
	\newblock Low-noise phase of a two-dimensional active nematic system.
	\newblock {\em Physical Review E}, 97(1):012707, 2018.
	
	\bibitem{tasaki2020hohenberg}
	Hal Tasaki.
	\newblock Hohenberg-mermin-wagner-type theorems for equilibrium models of
	flocking.
	\newblock {\em Physical Review Letters}, 125(22):220601, 2020.
	
	\bibitem{levis2017synchronization}
	Demian Levis, Ignacio Pagonabarraga, and Albert D{\'\i}az-Guilera.
	\newblock Synchronization in dynamical networks of locally coupled
	self-propelled oscillators.
	\newblock {\em Physical Review X}, 7(1):011028, 2017.
	
	\bibitem{gregoire2003moving}
	Guillaume Gr{\'e}goire, Hugues Chat{\'e}, and Yuhai Tu.
	\newblock Moving and staying together without a leader.
	\newblock {\em Physica D: Nonlinear Phenomena}, 181(3-4):157--170, 2003.
	
	\bibitem{theurkauff2012dynamic}
	Isaac Theurkauff, C{\'e}cile Cottin-Bizonne, J{\'e}r{\'e}mie Palacci,
	Christophe Ybert, and Lydric Bocquet.
	\newblock Dynamic clustering in active colloidal suspensions with chemical
	signaling.
	\newblock {\em Physical review letters}, 108(26):268303, 2012.
	
	\bibitem{buttinoni2013dynamical}
	Ivo Buttinoni, Julian Bialk{\'e}, Felix K{\"u}mmel, Hartmut L{\"o}wen, Clemens
	Bechinger, and Thomas Speck.
	\newblock Dynamical clustering and phase separation in suspensions of
	self-propelled colloidal particles.
	\newblock {\em Physical review letters}, 110(23):238301, 2013.
	
	\bibitem{palacci2013living}
	Jeremie Palacci, Stefano Sacanna, Asher~Preska Steinberg, David~J Pine, and
	Paul~M Chaikin.
	\newblock Living crystals of light-activated colloidal surfers.
	\newblock {\em Science}, 339(6122):936--940, 2013.
	
	\bibitem{fily2012athermal}
	Yaouen Fily and M~Cristina Marchetti.
	\newblock Athermal phase separation of self-propelled particles with no
	alignment.
	\newblock {\em Physical review letters}, 108(23):235702, 2012.
	
	\bibitem{bialke2013microscopic}
	Julian Bialk{\'e}, Hartmut L{\"o}wen, and Thomas Speck.
	\newblock Microscopic theory for the phase separation of self-propelled
	repulsive disks.
	\newblock {\em EPL (Europhysics Letters)}, 103(3):30008, 2013.
	
	\bibitem{redner2013structure}
	Gabriel~S Redner, Michael~F Hagan, and Aparna Baskaran.
	\newblock Structure and dynamics of a phase-separating active colloidal fluid.
	\newblock {\em Physical review letters}, 110(5):055701, 2013.
	
	\bibitem{bialke2012crystallization}
	Julian Bialk{\'e}, Thomas Speck, and Hartmut L{\"o}wen.
	\newblock Crystallization in a dense suspension of self-propelled particles.
	\newblock {\em Physical review letters}, 108(16):168301, 2012.
	
	\bibitem{maitra2019oriented}
	Ananyo Maitra and Sriram Ramaswamy.
	\newblock Oriented active solids.
	\newblock {\em Physical Review Letters}, 123(23):238001, 2019.
	
	\bibitem{xing2003thermal}
	Xiangjun Xing and Leo Radzihovsky.
	\newblock Thermal fluctuations and anomalous elasticity of homogeneous nematic
	elastomers.
	\newblock {\em EPL (Europhysics Letters)}, 61(6):769, 2003.
	
	\bibitem{xing2008nonlinear}
	Xiangjun Xing and Leo Radzihovsky.
	\newblock Nonlinear elasticity, fluctuations and heterogeneity of nematic
	elastomers.
	\newblock {\em Annals of Physics}, 323(1):105--203, 2008.
	
	\bibitem{warner2007liquid}
	Mark Warner and Eugene~Michael Terentjev.
	\newblock {\em Liquid crystal elastomers}, volume 120.
	\newblock Oxford university press, 2007.
	
	\bibitem{ferrante2013elasticity}
	Eliseo Ferrante, Ali~Emre Turgut, Marco Dorigo, and Cristi{\'a}n Huepe.
	\newblock Elasticity-based mechanism for the collective motion of
	self-propelled particles with springlike interactions: A model system for
	natural and artificial swarms.
	\newblock {\em Physical review letters}, 111(26):268302, 2013.
	
	\bibitem{ferrante2013collective}
	Eliseo Ferrante, Ali~Emre Turgut, Marco Dorigo, and Cristian Huepe.
	\newblock Collective motion dynamics of active solids and active crystals.
	\newblock {\em New Journal of Physics}, 15(9):095011, 2013.
	
	\bibitem{vicsek1995novel}
	Tam{\'a}s Vicsek, Andr{\'a}s Czir{\'o}k, Eshel Ben-Jacob, Inon Cohen, and Ofer
	Shochet.
	\newblock Novel type of phase transition in a system of self-driven particles.
	\newblock {\em Physical review letters}, 75(6):1226, 1995.
	
	\bibitem{huepe2015scale}
	Cristi{\'a}n Huepe, Eliseo Ferrante, Tom Wenseleers, and Ali~Emre Turgut.
	\newblock Scale-free correlations in flocking systems with position-based
	interactions.
	\newblock {\em Journal of Statistical Physics}, 158(3):549--562, 2015.
	
	\bibitem{kloeden2013numerical}
	Peter~E Kloeden and Eckhard Platen.
	\newblock {\em Numerical solution of stochastic differential equations},
	volume~23.
	\newblock Springer Science \& Business Media, 2013.
	
	\bibitem{li2019accurate}
	Yan-Wei Li and Massimo~Pica Ciamarra.
	\newblock Accurate determination of the translational correlation function of
	two-dimensional solids.
	\newblock {\em Physical Review E}, 100(6):062606, 2019.
	
	\bibitem{adhyapak2013live}
	Tapan~Chandra Adhyapak, Sriram Ramaswamy, and John Toner.
	\newblock Live soap: stability, order, and fluctuations in apolar active
	smectics.
	\newblock {\em Physical review letters}, 110(11):118102, 2013.
	
	\bibitem{chen2013universality}
	Leiming Chen, John Toner, et~al.
	\newblock Universality for moving stripes: A hydrodynamic theory of polar
	active smectics.
	\newblock {\em Physical review letters}, 111(8):088701, 2013.
	
	\bibitem{turgut2008self}
	Ali~E Turgut, Hande {\c{C}}elikkanat, Fatih G{\"o}k{\c{c}}e, and Erol
	{\c{S}}ahin.
	\newblock Self-organized flocking in mobile robot swarms.
	\newblock {\em Swarm Intelligence}, 2(2-4):97--120, 2008.
	
	\bibitem{ccelikkanat2010steering}
	Hande {\c{C}}elikkanat and Erol {\c{S}}ahin.
	\newblock Steering self-organized robot flocks through externally guided
	individuals.
	\newblock {\em Neural Computing and Applications}, 19(6):849--865, 2010.
	
	\bibitem{ferrante2010flocking}
	Eliseo Ferrante, Ali~Emre Turgut, Nithin Mathews, Mauro Birattari, and Marco
	Dorigo.
	\newblock Flocking in stationary and non-stationary environments: a novel
	communication strategy for heading alignment.
	\newblock In {\em International conference on parallel problem solving from
		nature}, pages 331--340. Springer, 2010.
	
	\bibitem{van2019interrupted}
	Marjolein~N Van Der~Linden, Lachlan~C Alexander, Dirk~GAL Aarts, and Olivier
	Dauchot.
	\newblock Interrupted motility induced phase separation in aligning active
	colloids.
	\newblock {\em Physical review letters}, 123(9):098001, 2019.
	
	\bibitem{lavergne2019group}
	Fran{\c{c}}ois~A Lavergne, Hugo Wendehenne, Tobias B{\"a}uerle, and Clemens
	Bechinger.
	\newblock Group formation and cohesion of active particles with visual
	perception--dependent motility.
	\newblock {\em Science}, 364(6435):70--74, 2019.
	
\end{thebibliography}
\end{document}


	\title{Supplementary Information\\
		Alignment Destabilizes Crystal Orders in Active Systems}
	\author{Chen Huang$^{1}$}
	\author{Leiming Chen$^{2}$}
	\author{Xiangjun Xing$^{1,3,4}$}
	\email{xxing@sjtu.edu.cn}
	\address{$^1$ Wilczek Quantum Center, School of Physics and Astronomy, Shanghai Jiao Tong University, Shanghai 200240 China\\
$^2$ School of Materials Science and Physics,
China University of Mining and Technology, Xuzhou, Jiangsu, 221116  China\\
	$^3$ T.D. Lee Institute, Shanghai Jiao Tong University, Shanghai 200240 China \\
	$^4$ Shanghai Research Center for Quantum Sciences, Shanghai 201315 China}
	
	
	

\maketitle

\newpage

\begin{widetext}

\vspace{5mm}
\section{Dimensionless form of the equations of motion and the choice of parameters}
We derive in this section the dimensionless equations of the model and specify the parameters used in simulation. To distinguish the coefficients used in simulation with their counterparts in analytical results, we temporarily add a subscript or superscript $L$ (meaning ``Lattice") in this section. 

In the simulation model, the space and time units are chosen as $a_0$ (lattice constant) and $\tau_0$ (the value of $\tau_0$ would be determined later). The dimensionless control parameters are: 
\ba
&&\tilde{b}_L\equiv b\tau_0/(\gamma a_0),\quad \tilde{\kappa}\equiv\kappa\tau_0/\gamma, \quad \tilde{D}^L\equiv D\tau_0/a_0^2, \quad \tilde{D}_{\theta}^L\equiv D_{\theta}\tau_0, \quad \tilde{d}_L\equiv d\tau_0, \quad \tilde{c}_L\equiv c\gamma a_0, \nonumber\\
&&\widetilde{\mathbf{F}}_k\equiv\mathbf{F}_k\tau_0/(\gamma a_0),\quad \widetilde{\eta}_k\equiv\eta\sqrt{\tau_0}, \quad \tilde{\hat{\xi}}_k\equiv\hat{\xi}\sqrt{\tau_0}. 
\ea
The dimensionless equation of Eq.~(1) is,
\be
\dot{ \tilde{ \mathbf{r} }}_k( \tilde{t} ) 
= 
\tilde{b}_L \hat{n}(\theta_k) + \widetilde{\mathbf{F}}_k(\tilde{t}) + \sqrt{2\tilde{D}^L} \tilde{\hat{\xi}}_k(\tilde{t}),
\ee
The dimensionless equations of the two alignment dynamics become:
\ba
&&\dot{\theta}_k(\tilde{t}) = \tilde{d}_L ( \hat{n}(\theta_k) \times \langle\hat{n} \rangle_k) \cdot \hat{z} + \sqrt{2\tilde{D}_{\theta}^L} \tilde{\eta}_k,\quad
\text{(AD-$\widetilde{\text{I}}$)}\\
&&\dot{\theta}_k(\tilde{t}) = \tilde{c}_L ( \hat{n}(\theta_k) \times \widetilde{\mathbf{F}}_k ) \cdot \hat{z} + \sqrt{2\tilde{D}_{\theta}^L} \tilde{\eta}_k,\quad
\text{(AD-$\widetilde{\text{II}}$)}
\ea

We numerically integrate the above dynamic equations using the first-order Euler-Maruyama scheme~\cite{kloeden2013numerical}. We choose $\tau_0\equiv 0.01 a_0^2/D^L$ to ensure the perturbation assumption such that the magnitude of the noise $\sqrt{\tilde{D}^L}=0.1$ is small compared to the active force and the elastic force. The other parameters are chosen as:
\ba
\tilde{b}_L = 2,\ \tilde{\kappa}=100 \text{ for AD-I, } \tilde{\kappa}=20 \text{ for AD-II}.
\ea


In determining the phase diagram, we use a relatively small lattice of size $32\times 32$ to save computational resources, start from a perfect triangular lattice with all active forces orienting randomly.

In studying the translational order and other correlation functions, we use a larger lattice of size $256\times 256$. For the stationary phase, we choose the parameters $\tilde{D}_{\theta}^L = 0.3$, $\tilde{d}=0.1$ for both AD-I and AD-II, and start from a perfect triangular lattice and all active forces orienting randomly. For the collective moving phase, we use the parameter $\tilde{D}_{\theta}^L = 0.05$ (Note that by our choice, the magnitude of the angular noise also satisfies the small perturbation condition $\sqrt{\tilde{D}_{\theta}^L}<1$.), $\tilde{d}=10$ for AD-I and $\tilde{d}=1$ for AD-II, and also start from a perfect lattice, but with all active forces orienting toward the $\xh$ axis. Note that for any finite-size lattice, the average velocity of the collectively moving crystal would slowly change, so we redirect it back toward the $\xh$ axis right after each sample has been collected but before the subsequent simulation steps start.

In the coarse-grained continuous model, using the space and time units $a_0$ and $\tau_0$, we obtain the reduced quantities:
\ba
&& \tilde{b} \equiv \frac{b \tau_0}{\gamma a_0 },\ 
\widetilde{\mathbf{F}} \equiv \frac{\mathbf{F} \tau_0}{\gamma a_0},\ 
\left(\begin{matrix}\tilde{\mu}\\ \tilde{\lambda}\end{matrix}\right)\equiv \frac{\tau_0}{\gamma a_0^2}\left(\begin{matrix}\mu\\ \lambda\end{matrix}\right),\ 
\tilde{D}\equiv \frac{D\tau_0}{a_0^2}, \ 
\tilde{D}_\theta\equiv D_\theta\tau_0,\nonumber \\ 
&& \tilde{d}\equiv\frac{d\tau_0}{a_0^2}, \ 
\tilde{c}\equiv c\gamma a_0,\ 
\left(\begin{matrix}\tilde{\eta}\\ \tilde{\xi}\end{matrix}\right)\equiv \sqrt{\tau_0}\left(\begin{matrix}\eta\\ \xi\end{matrix}\right).
\ea
The dimensionless form of Eq.~(4) is:
\ba
\dot{\tilde{\mathbf{u}}}(\tilde{\mathbf{r}},\tilde{t}) =
\tilde{b} \theta \hat{y} +
\widetilde{\mathbf{F}}(\tilde{\mathbf{r}},\tilde{t}) +
\sqrt{2\tilde{D}} \hat{\xi}(\tilde{\mathbf{r}},\tilde{t}).
\ea
The two coarse-grained dimensionless alignment dynamics are, respectively,
\ba
&&\dot{\theta}(\tilde{\mathbf{r}},\tilde{t}) =
\tilde{d}\tilde{\Delta}\theta +
\sqrt{2\tilde{D}_\theta} \tilde{\eta}(\tilde{\mathbf{r}},\tilde{t}),\quad\text{(AD-$\tilde{\text{I}}'$)}\\
&&\dot{\theta}(\tilde{\mathbf{r}},\tilde{t}) =
\tilde{c}(\hat{n}\times\widetilde{\mathbf{F}})\cdot\hat{z} + 
\sqrt{2\tilde{D}_\theta} \tilde{\eta}(\tilde{\mathbf{r}},\tilde{t}).\quad\text{(AD-$\tilde{\text{II}}'$)}
\ea
where $\tilde{\Delta}\equiv a_0^2\Delta$. Since the linearized elasticity theory in the continuous model also applies to the two-dimensional hexagonal crystal used in our simulation~\cite{chaikin1995principles}, we may relate the $Lam\acute{e}$ constants and the elastic constant via $\tilde{\lambda}=\tilde{\mu}=\tilde{\kappa}\sqrt{3}/4$~\cite{hoover1974two} in order to compare the two models. 

\section{Fourier Transform on non-orthogonal lattice}
The simulations use a triangular lattice with rhomb-like boundary, where the lattice points are shown in Fig.~\ref{fig:nonOrthogonalLattice}(a) as the solid black points. We wish to implement the Fourier Transform (FT) to the data (e.g., the displacement $u_x$, $u_y$, and the angle $\theta$) sampled on this triangular lattice into the reciprocal $k$-space. Although it is easy to implement FT on a square lattice by using the Fast Fourier Transform (FFT) techniques, no direct method exists for Ft on a triangular lattice. However, the triangular lattice can be deemed as originating from a linear transform of the square lattice. Since FT is also a linear transform, we expect a combination of the two linear transforms may serve our purpose. Below is a derivation of this process.

Suppose we sample a complex function $f : \mathbb{R}^{n} \rightarrow \mathbb{C}$ on a set of n-dimensional points $\Lambda\in\mathbb{R}^n$ in real space. Let $\Phi:\mathbb{R}^n\rightarrow\mathbb{R}^n$ be a linear isomorphic map under which n-dimensional integers are mapped to n-dimensional real values such that $\mathbb{Z}^{n} \rightarrow \Lambda$. We may obtain the FT of $f$ in terms of the FT of $(f \circ \Phi)$, which is a function on $\mathbb{Z}^{n}$. The FT of $(f \circ \Phi)$ is,
\ba
\mathscr{F}(f \circ \Phi)(s) &\equiv& 
\int f \circ \Phi(x) e^{-2 \pi i\langle s, x\rangle} d x\\
&=&
\frac{1}{|\operatorname{det} \Phi|} \int f(y) e^{-2 \pi i\langle\Phi^{-T}s, y\rangle} dy\nonumber\\
&=&
\frac{1}{|\operatorname{det} \Phi|} \mathscr{F}(f) \left(\Phi^{-T} s\right)
\ea
Here a substitution $y=\Phi x$ was made, and it was used that $\left\langle s, \Phi^{-1} y\right\rangle=\left\langle\Phi^{-T} s, y\right\rangle .$ Note that the conventions with respect to the sign of the exponent and factors of $2 \pi$ may differ.

We now discuss the meaning of the above formula. The function $f \circ \Phi$ is the value of the sample points of $f$ arranged in an $n$-dimensional array lying on a \textit{standard grid} (equal-axis orthogonal grid). Its Fourier transform gives an array of the same shape, but the interpretation is different: the value at index $s$ in reality is the value at $\Phi^{-T} s\in\mathbb{R}^{n}$ in $k$-space, differed by a multiplier $|\operatorname{det} \Phi|$.

In our simulation's setup, $n=2$ and $\Lambda\in\mathbb{R}^2$ is the set of the triangular lattice points in real space. The function $f$ whose FT we want to compute, can be the displacement field $u_x$, $u_y$ or the orientation field $\theta$. The linear map $\mathbf{y}=\Phi \mathbf{x}$ with $\Phi=\left(\begin{matrix}1 & \frac{1}{2}\\0&\frac{\sqrt{3}}{2}\end{matrix}\right)$ transforms a square regular lattice with lattice constant $a_0$ into the desired triangular lattice with the same lattice constant. The transverse of its inverse is $\Phi^{-T}=\left(\begin{matrix}1 & 0\\-\frac{1}{\sqrt{3}}&\frac{2}{\sqrt{3}}\end{matrix}\right)$. The function $f\circ\Phi$ takes the value of the sample points of $f$ on $\Lambda$, but its arguments lie on the corresponding square lattice points. We apply FFT on this real space square lattice (standard grid), obtain its FT values on a $k$-space square lattice (standard grid) indexed by $s$, and assign the FT values to its true coordinate in $k$-space by left multiplying $\Phi^{-T}$ to $s$. Suppose the standard grid in real space has a linear size $L\times L$ and each dimension is divided into equally spaced $n$ points with distance of unity $a_0=1$, then there are $n\times n$ total sample points in both real and $k$-space.

\begin{figure}[ht!]
	\centering
	\includegraphics[width=6in]{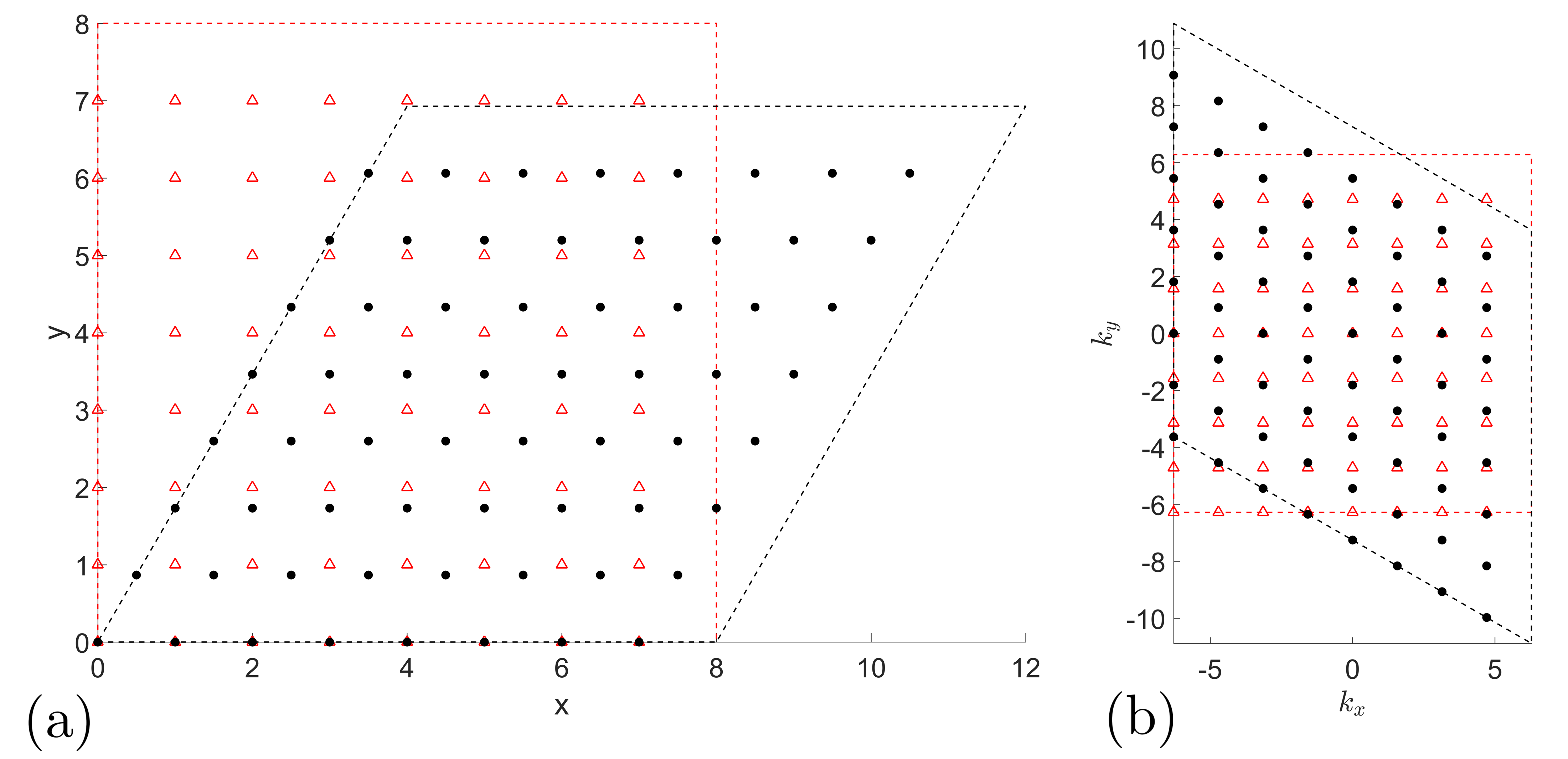}
	\caption{A lattice with lattice constant $a_0=1$ and $L=8$. The black points denote (a) Sample points in real space and (b) the corresponding k-space points. The red triangles are the standard grid in each space respectively. The dashed lines denote the boundary of the simulation box.}
	\label{fig:nonOrthogonalLattice}
\end{figure}

In Fig.~\ref{fig:nonOrthogonalLattice}, a lattice of $L=8$ and $n=8$ is shown. The lattice constant of the standard grid in real space is $a_0=1$ and that in k-space is $\Delta k=2\pi/L=\pi/4$. The k-space vector of the standard grid with the minimal magnitude $\mathbf{s_1}=(0,\frac{2\pi}{L})$ transforms to the true k-space coordinate, $\mathbf{g_1}=(0,\frac{2\pi}{L}\frac{2}{\sqrt{3}})$, while $\mathbf{s_2}=(\frac{2\pi}{L},0)$ transforms to $\mathbf{g_2}=(\frac{2\pi}{L},-\frac{1}{\sqrt{3}}\frac{2\pi}{L})$ with both $|\mathbf{g_i}|=\frac{2\pi}{L}\frac{2}{\sqrt{3}}$.

\section{Translational Correlation, Pair Correlation, and Bond-Orientational Correlation}
In this section, we discuss the definition and evaluation of the translational correlation, pair correlation of velocity and director of active force, and bond-orientational correlation in AD-I and AD-II dynamics. For simplicity, the quantities such as $r$ and $\mathbf{q}$ in this section all refer to dimensionless variables. 

We first follow the steps in Ref.~\cite{li2019accurate} by defining the translational correlation function as:
\ba
g_{\mathbf{q}}(r) = \frac{1}{2 \pi r \Delta r \rho N} \sum_{j \neq k} \zeta\left(r - \left|\mathbf{r}_{j} - \mathbf{r}_{k}\right|\right) \text{Re} ( e^{i \mathbf{q} \cdot\left(\mathbf{r}_{j} - \mathbf{r}_{k}\right)} ).
\ea
where $\zeta=1$ if $\left|\mathbf{r}_{j}-\mathbf{r}_{k}\right|$ is in the region $r\sim r+\Delta r$, otherwise $\zeta=0$. The number density $\rho = \left(\sqrt{3}/2\right)^{-1}$ is the inverse of the area of a unit cell in the triangular lattice of unit length. We choose $\mathbf{q}$ by evaluating the Bragg peaks of the structure factor
\ba
Q(q_x,q_y) = \frac{1}{N}\langle \rho(q_x,q_y) \rho(-q_x,-q_y) \rangle,
\ea
where
\ba
\rho\left(q_{x}, q_{y}\right) = \sum_{j=1}^{N}  \exp \left[i\left(q_{x} x_{j} + q_{y} y_{j}\right)\right],
\ea
with $x_j$ and $y_j$ the Cartesian coordinates of the $j$th particle. The choice of $\mathbf{q}$ should be carefully decided from simulation results, i.e., the true peak value is numerically evaluated around the Bragg peak of a perfect hexagonal lattice $\mathbf{q}_x= (4\pi,0)$ along the $\hat{x}$ axis, $\mathbf{q}_y=(0,2\pi\cdot\left(\sqrt{3}/2\right)^{-1})$ along the $\hat{y}$ axis, and $\mathbf{q}_t= 2\pi(1,1/\sqrt{3})$ along a tilted direction. 

In Fig.~\ref{fig::BraggPeakADI} and Fig.~\ref{fig::BraggPeakADII}, we plot $Q(q_x,q_y)$ (the first column) and the corresponding translational correlation function (the second column in Log-Log scale and the third in Log-Linear scale) of typical sample configurations, for AD-I with $(\tilde{\kappa},\tilde{d}_L)=(100,10)$ and AD-II with $(\tilde{\kappa},\tilde{c}_L)=(20,1)$, respectively. Data is collected after the system has reached a steady state of collective moving, with $P\approx 0.9915$ in AD-I and $P\approx 0.9830$ in AD-II. In AD-I, the translational correlation function decays faster than power law along the wave vector $\mathbf{q}_x$, $\mathbf{q}_y$ and $\mathbf{q}_t$, indicating short-range order. However, in AD-II, it has a power law decay along the $\hat{x}$ axis, and exhibits short-range order along $\mathbf{q}_y$ and $\mathbf{q}_t$.

\begin{figure*}[t!]
	\centering
	\includegraphics[height=5in]{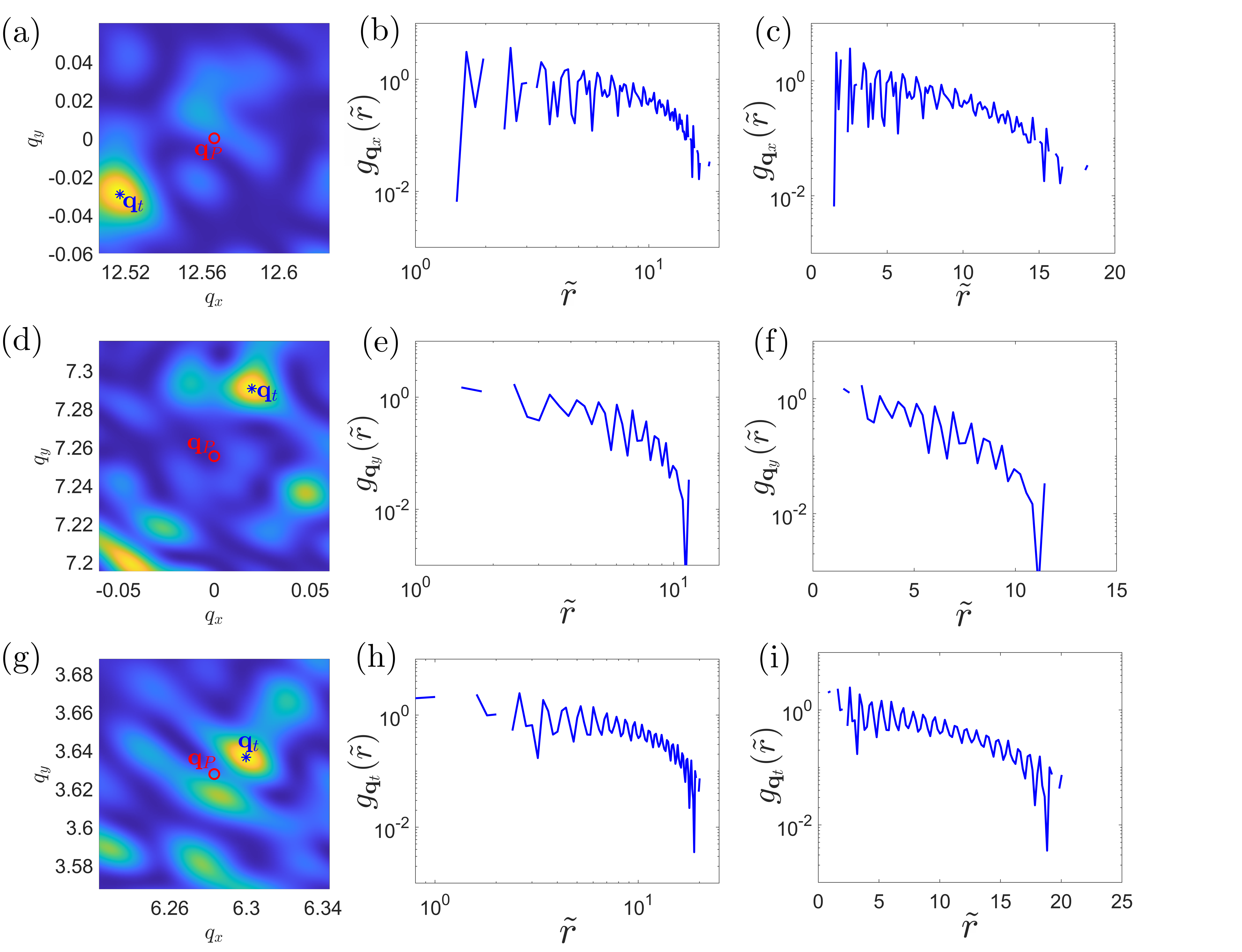}
	\caption{Bragg peak of $Q(q_x,q_y)$ and the corresponding translational correlation function in AD-I with (a)-(c) $\mathbf{q}_x=(12.5172,-0.0292)$; (d)-(f) $\mathbf{q}_y=(0.0196,7.2905)$; (g)-(i) $\mathbf{q}_t=(6.2998,3.6363)$. The second column is in Log-Log scale, while the third column is in Log-Linear scale.}
	\label{fig::BraggPeakADI}
\end{figure*}

\begin{figure*}[t!]
	\centering
	\includegraphics[height=5in]{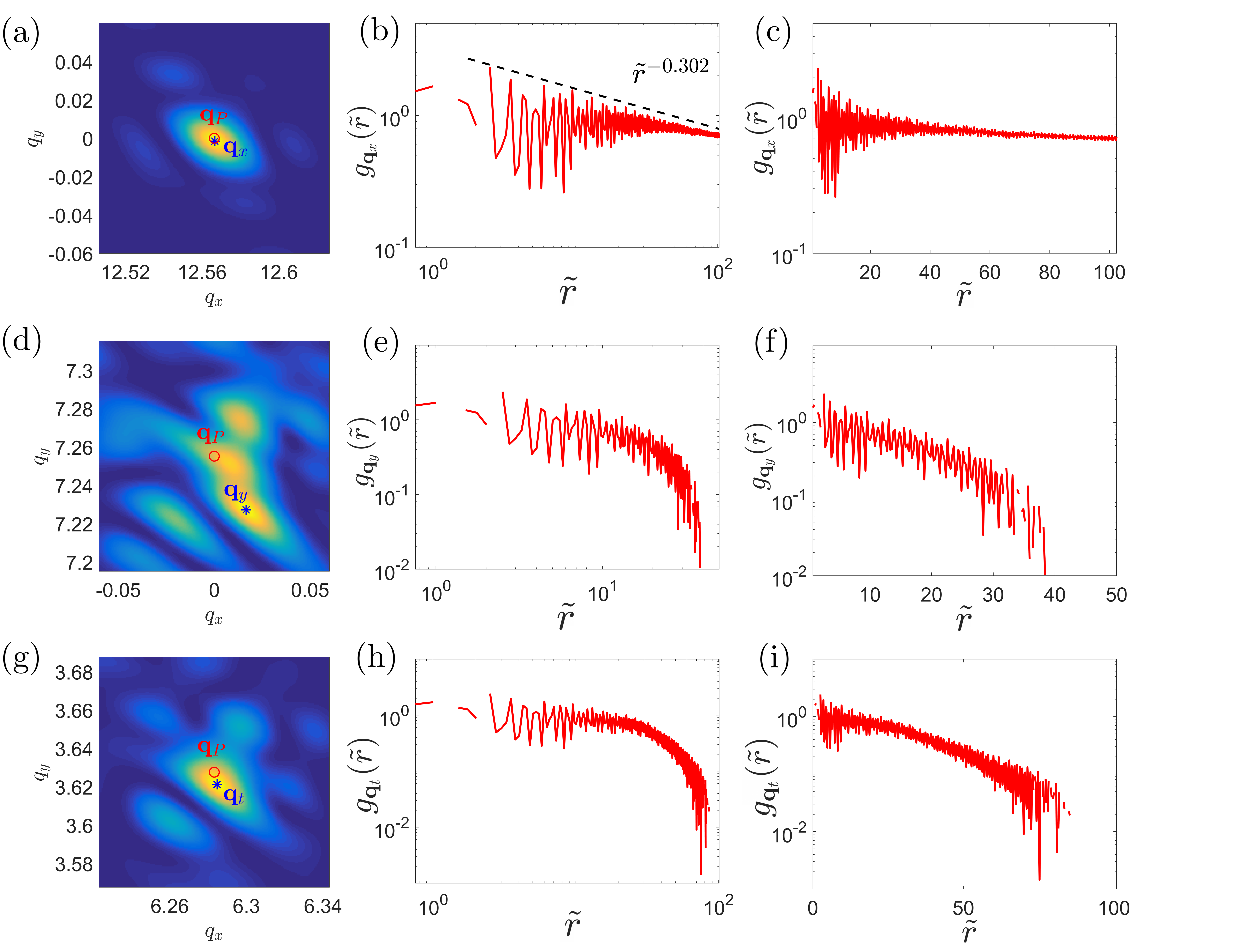}
	\caption{Bragg peak of $Q(q_x,q_y)$ and the corresponding translational correlation function in AD-II with (a)-(c) $\mathbf{q}_x=(12.5667,-0.0015)$; (d)-(f) $\mathbf{q}_y=(0.0166,7.2272)$; (g)-(i) $\mathbf{q}_t=(6.2847,3.6213)$. The second column is in Log-Log scale, while the third column is in Log-Linear scale.}
	\label{fig::BraggPeakADII}
\end{figure*}

In the first row of Fig.~\ref{fig::bondOrder}, we show the correlation function of the director of active force in k-space $\langle|\hat{\theta}|^2\rangle$. We observe a $k^{-2}$ decay in AD-I, and a $k^{-2}$ decay followed by a constant in AD-II. These scaling properties match with the analytical results Eqn.~(\ref{S_theta_ADI}) and (\ref{S_theta_ADII}) given in Appendix D.

In the second row of Fig.~\ref{fig::bondOrder}, we plot 10 independent samples at different times of the the pair correlation function of the director of active force,
\ba
\langle \hat{n}(0)\cdot\hat{n}(r)\rangle = \frac{1}{2 \pi r \Delta r \rho N} \sum_{j \neq k} \zeta\left(r - \left|\mathbf{r}_{j} - \mathbf{r}_{k}\right|\right) (\hat{n}_{j} \cdot \hat{n}_{k}).
\ea
Note we have subtracted the asymptotic value of $\langle \hat{n}(0)\cdot\hat{n}(r)\rangle$ at large enough distance $\langle\hat{n}(0)\cdot\hat{n}(\infty)\rangle$ in the steady state, i.e., $\langle \hat{n}(0)\cdot\hat{n}(r)\rangle_c \equiv \langle \hat{n}(0)\cdot\hat{n}(r)\rangle - \langle\hat{n}(0)\cdot\hat{n}(\infty)\rangle $. We find a power-law decay with exponent being approximately $-1$ indicating quasi-long range order. A similar treatment is applied to the velocity with
\ba
\langle \hat{v}(0)\cdot\hat{v}(r)\rangle = \frac{1}{2 \pi r \Delta r \rho N} \sum_{j \neq k} \zeta\left(r - \left|\mathbf{r}_{j} - \mathbf{r}_{k}\right|\right) (\vh_{j} \cdot \vh_{k}).
\ea
where the velocity $\mathbf{v}_i$ of each particle is defined by subtracting the coordinates in the two configurations separated by time interval $100\Delta \tilde{t}$, i.e., $\mathbf{v}_i(\tilde{t})\equiv [\mathbf{r}_i(\tilde{t})-\mathbf{r}_i(\tilde{t}-100\Delta \tilde{t})]/(100\Delta \tilde{t})$ and $\hat{v}_i \equiv \mathbf{v}_i/|\mathbf{v}_i|$. We also define and plot the value $\langle \hat{v}(0)\cdot\hat{v}(r)\rangle_c \equiv \langle \hat{v}(0)\cdot\hat{v}(r)\rangle - \langle\hat{v}(0)\cdot\hat{v}(\infty)\rangle $ as shown in Fig.~3(e)-(f) in the main text.
The result gives a power-law decay of similar exponents with the director of active force.

In Fig.~\ref{fig::bondOrder}(e)-(f), we show the bond-orientational correlation function~\cite{li2019accurate}:
\ba
g_{6}(r) = \text{Re}\left\langle \psi_{6}\left(\vec{r}_{i}\right)  \psi_{6}^{*}\left(\vec{r}_{j}\right) \right\rangle
\equiv \frac{1}{2 \pi r \Delta r \rho N} \sum_{j \neq k} \zeta\left(r - \left|\mathbf{r}_{j} - \mathbf{r}_{k}\right|\right)
\text{Re}( \psi_6\left(\vec{r}_{i}\right)  \psi_6^*\left(\vec{r}_{j}\right) ),
\ea
where $r = |\vec{r}_i-\vec{r}_j|$. The local-bond orientational parameter is defined as $\psi_{6}\left(\vec{r}_{j}\right)=\frac{1}{n} \sum_{m=1}^{n} \exp i (6 \theta_{m}^{j})$ where the sum runs over the $n$ Voronoi neighbors of the particle, and $\theta_m^j$ is the angle of $(\vec{r}_m-\vec{r}_j)$ relative to any fixed axis. In Fig.~\ref{fig::bondOrder}(e)-(f), $g_6(r)$ in AD-I has a power law decay with a non-zero but small exponent $-0.0284$. This exponent is obtained from linear fitting of the average of 10 independent sample curves at different simulation steps, after the system has reached the steady state. The value of $g_6(r)$ in AD-II approaches a constant at large distances. From the numerical results we may conclude that the bond-orientational order in AD-I is QLRO and that in AD-II is LRO. We leave the verification of these predictions and the evaluation of a more accurate power-law exponent to future studies.

\begin{figure*}[t!]
	\centering
	\includegraphics[width=6in]{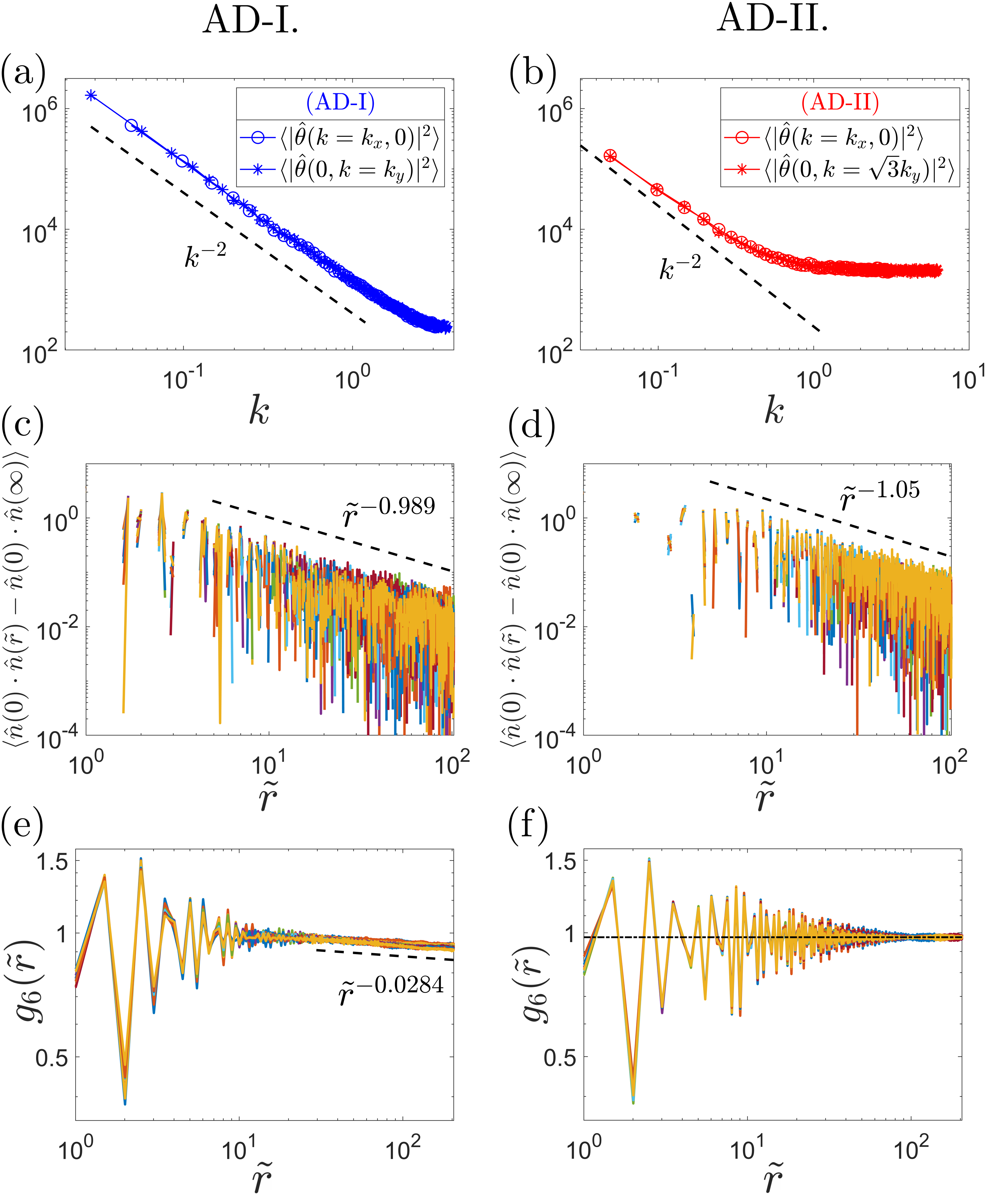}
	\caption{(a)-(b) The correlation functions $\langle|\hat{\theta}|\rangle$ along $\hat{k}_x$ and $\hat{k}_y$ axis (Note that the $k_y$ value in AD-II is rescaled for comparison according to Eq.~(\ref{S_theta_ADII}).), (c)-(d) the pair correlation functions of director of active force, and (e)-(f) the bond-orientational correlation functions of AD-I (the first column) and AD-II (the second column).}
	\label{fig::bondOrder}
\end{figure*}

\section{Analytical Solution of the correlation functions}

To obtain a more thorough understanding of these numerical results, and also to understand analogous systems in higher dimensions, we resort to analytical approach.  Our lattice model of active crystal can be coarse-grained to yield a continuous theory that is amenable for analytic study.  Let $\rv$ be the Lagrange coordinate for a generic swimmer which denotes its position in the initial reference state, $\mathbf{u}(\mathbf{r},t)$ the phonon field characterizing its displacement from the uniformly moving state, the swimmer's position at time $t$ is then $\rv + \uv(\rv,t) + {\bf v}_0 t$, where ${\bf v}_0 = b/\gamma$ is the mean velocity as determined from Eq.~(1) in the main text (Hence our analytic results apply to the moving phase). The elastic force $\mathbf{F}(\mathbf{r},t)$ acting on the swimmer at $\rv$ can be expressed in terms of $\uv(\rv,t)$ using the linearized elasticity theory~\cite{landau1986theory,fetter2003theoretical}:   
\begin{subequations}
	\ba
	\mathbf{F} 
	= \mu\nabla^2\mathbf{u} 
	+ (\lambda+\mu) 
	\nabla(\nabla\cdot\mathbf{u}),  
	\label{eq:elasticforce}
	\ea
	where $\lambda$ and $\mu $ are two $Lam\acute{e}$ coefficients.  This result is the coarse-grained version of Eq.~(2) in the main text. In terms of components we have
	\ba
	\left\{\begin{array}{ll}
		F_{x}&=(\lambda+2 \mu) \partial_x^{2} u_{x} + \mu \partial_y^{2} u_{x} + (\lambda+\mu) {\partial_x \partial_y u_{y}}, 
		\vspace{2mm}
		\\
		F_{y}&=(\lambda+2 \mu) {\partial^{2}_y u_{y}} + \mu {\partial^{2}_x u_{y}}
		+ (\lambda+\mu) {\partial_x \partial_y u_{x}}.
	\end{array}\label{eq:explicitElasticForce}
	\right.
	\ea
\end{subequations}



The Langevin equations are coarse-grained into stochastic PDEs, which can be further linearized.  In particular Eq.~(1) in the main text becomes
\ba
\gamma \,  \dot{\mathbf{u}}(\mathbf{r},t)
= b \, \theta (\mathbf{r},t)\, \hat{y} + \mathbf{F} (\mathbf{r},t) 
+ \gamma \, \sqrt{2D} \,\hat{\xi}(\mathbf{r},t). 
\label{eq:overdamped}
\ea
where $\theta (\mathbf{r},t)$ is the coarse-grained version of $\theta_k(t)$, and $\hat{\xi}(\mathbf{r},t)$ is a 2d Gaussian white noise satisfying $\langle\xi_i(\mathbf{r},t)\rangle=0,\ \langle\xi_i(\mathbf{r},t)\xi_j(\mathbf{r}',t')\rangle=\delta_{ij}\delta(\mathbf{r}-\mathbf{r}')\delta(t-t')$. The stochastic PDE for the phonon field can be obtained by substituting Eq.~(\ref{eq:explicitElasticForce}) into Eq.~(\ref{eq:overdamped}).  Here in Eq.~(\ref{eq:overdamped}) and also below in Eqs.~(\ref{eq:peturb_model1-0}), (\ref{eq:peturb_model2-0}), the parameters $b, c,d, \gamma, D, D_{\theta}$ are in general different from their lattice versions in Eqs.~(1), (AD-I) and (AD-II) in the main text.  We however need not to be concerned with these differences since we are only interested in the scaling behaviors of correlation functions in the analytic theory. 

Coarse-graining and linearizing Eq.~(AD-I), we obtain the following set of linearized stochastic PDEs for AD-I:
\ba
\left\{\begin{array}{ll}
	\gamma \, \dot{u}_x&=F_x+\sqrt{2\gamma T} \, \xi_x\\
	\gamma \, \dot{u}_y&=b \, \theta+F_y+\gamma\sqrt{2D}\, \xi_y\\
	\dot{\theta} &=d \,\Delta\theta+\sqrt{2D_\theta}\, \eta
	\label{eq:peturb_model1-0}
\end{array}.
\right.
\ea
where $\eta= \eta(\mathbf{r},t)$ is a Gaussian white noise satisfying $\langle\eta(\mathbf{r},t)\rangle=0,\ \langle\eta(\mathbf{r},t)\eta(\mathbf{r}',t')\rangle=\delta(\mathbf{r}-\mathbf{r}')\delta(t-t')$.   Further carrying out the spatial Fourier transform of Eqs.~(\ref{eq:peturb_model1-0}), we obtain linear Langevin equations for ${\mathbf{u}}(\kv, t)$ and $\theta(\kv, t)$.

The linearized theory of AD-II can be similarly obtained:
\ba
\left\{\begin{array}{ll}
	\gamma \, \dot{u}_x&= F_x+\gamma\sqrt{2D}\, \xi_x\\
	\gamma \, \dot{u}_y&= b \, \theta+F_y+\gamma\sqrt{2D} \, \xi_y\\
	\dot{\theta}&= c\, F_y + \sqrt{2D_\theta}\, \eta
\end{array}.
\right.
\label{eq:peturb_model2-0}
\ea

We are now ready to derive the steady state correlation function solution for ${\mathbf{u}}(\kv, t)$ and $\theta(\kv, t)$. But before that, let us first consider the solution of a general multidimensional Langevin equation:
\ba
\dot{\mathbf{x}}(\kv,t) + 
\mathbf{\Gamma} \mathbf{x}(\kv,t) =
\zeta(\kv,t).
\ea
where $\langle \zeta_i(\kv,t)\zeta_j^\dagger(\kv',t') \rangle = \langle \zeta_i(\kv,t)\zeta_j(-\kv',t')\rangle = 2B_{ij}\delta(\kv-\kv')\delta(t-t')$. The matrices $\mathbf{B}$ and $\mathbf{\Gamma}$ are real, and $\mathbf{x}(\kv,t)$ and $\zeta(\kv,t)$ are column vectors. Define $\mathbf{y} = e^{\mathbf{\Gamma} t}\mathbf{x}$, it is easy to prove $d\mathbf{y}/dt = e^{\mathbf{\Gamma} t}\zeta$, and we have the formal solution:
\ba
\mathbf{x}(\kv,t) =
\int_{-\infty}^t e^{-\mathbf{\Gamma}(t-\tau)}\zeta(\kv,\tau) d\tau.
\ea
The correlation matrix
\ba
\langle\mathbf{x}(\kv,t)\mathbf{x}^\dagger(\kv',t')\rangle
&=&
\int_{-\infty}^t\int_{-\infty}^{t'} e^{-\mathbf{\Gamma}(t-\tau)}\ 2\mathbf{B}\ \delta(\kv-\kv')\delta(\tau-\tau')\ e^{-\mathbf{\Gamma}^\dagger(t'-\tau')}\ d\tau\ d\tau'\\
&=&
\int_{-\infty}^{\min(t,t')} e^{-\mathbf{\Gamma}(t-\tau)}\ 2\mathbf{B}\ \delta(\kv-\kv')\ e^{-\mathbf{\Gamma}^T(t'-\tau)}\ d\tau
\ea
where we used $\mathbf{\Gamma}^\dagger = \mathbf{\Gamma}^T$ in the last equation. Define the autocorrelation matrix as
\ba
\mathbf{M}\equiv\langle \mathbf{x}(\kv,t)\mathbf{x}^\dagger(\kv,t)\rangle=\int_{-\infty}^t e^{-\mathbf{\Gamma}(t-\tau)}2\mathbf{B}e^{-\mathbf{\Gamma}^T(t-\tau)}d\tau
\ea
we have
\ba
\mathbf{\Gamma} \mathbf{M} +
\mathbf{M}\mathbf{\Gamma}^T
&=&
\int_{-\infty}^t e^{-\mathbf{\Gamma}(t-\tau)}\ 2(\mathbf{\Gamma} \mathbf{B} + \mathbf{B}\mathbf{\Gamma}^T)\ e^{-\mathbf{\Gamma}^T(t-\tau)}\ d\tau\\
&=&
2\ e^{-\mathbf{\Gamma}t}\left[\int_{-\infty}^t \frac{d}{d\tau}\left(e^{\mathbf{\Gamma}\tau}\mathbf{B} e^{\mathbf{\Gamma}^T \tau}\right)\ d\tau\right] e^{-\mathbf{\Gamma}t}=2\mathbf{B}
\ea
This is a \textit{Lyapunov equation}~\cite{lyapunov1992general} that can be solved analytically.

In AD-I, after implementing Fourier transform for the spatial component, we obtain the Lyapunov equation:
\ba
\left(\begin{matrix} \dot{\hat{u}}_x \\ \dot{\hat{u}}_y\\ \dot{\hat{\theta}}
\end{matrix}\right)
+
\mathbf{\Gamma}\left(\begin{matrix} \hat{u}_x\\ \hat{u}_y\\ \hat{\theta}
\end{matrix}\right)
=
\left(\begin{matrix}\sqrt{2D}\hat{\xi}_x\\ \sqrt{2D}\hat{\xi}_y\\ \sqrt{2D_\theta}\hat{\eta}
\end{matrix}\right)
\ea
where
\ba
\mathbf{\Gamma} =
\frac{1}{\gamma}\left(\begin{matrix} \lambda(3k_x^2+k_y^2) & 2\lambda k_xk_y & 0 \\ 2\lambda k_xk_y & \lambda(3k_y^2+k_x^2) & -b \\ 0 & 0 & \gamma d(k_x^2+k_y^2)
\end{matrix}\right)
\ea
and
\ba
\mathbf{M} =
\left(\begin{matrix}\langle |\hat{u}_x|^2\rangle & \langle \hat{u}_x\hat{u}_y^*\rangle & \langle \hat{u}_x\hat{\theta}^* \rangle \\
	\langle \hat{u}_y\hat{u}_x^*\rangle & \langle |\hat{u}_y|^2\rangle & \langle \hat{u}_y\hat{\theta}^* \rangle\\
	\langle \hat{\theta}\hat{u}_x^*\rangle & \langle \hat{\theta}\hat{u}_y^* \rangle & \langle |\hat{\theta}_x|^2\rangle\end{matrix}\right),\
\mathbf{B}=\left(\begin{matrix}D & 0 & 0\\ 0 & D & 0\\0 & 0 & D_\theta\end{matrix}\right)
\ea
The solution is
\ba
\left\{
\begin{aligned}
	\langle |\hat{u}_x(\kv)|^2\rangle &= \frac{D (k_x^2 + 3k_y^2) \gamma }{ 3 k^4 \lambda} 
	+ \frac{b^2 D_\theta k_x^2 k_y^2 (d \gamma + 4\lambda) }{ 3k^{10} \lambda d(d^2 \gamma^2 + 4d\gamma\lambda + 3\lambda^2)}\\
	&\equiv \frac{W_1(\alpha)}{k^6} + \frac{W_2(\alpha)}{k^2},\nonumber\\
	\langle |\hat{u}_y(\kv)|^2\rangle &= \frac{D (3k_x^2 + k_y^2)\gamma}{3k^4 \lambda} 
	+ \frac{b^2D_\theta\left[d(3k_x^4 + 3k_x^2 k_y^2 + k_y^4)\gamma + (3k_x^2 + k_y^2)^2\lambda\right]}{3k^{10}\lambda d(d^2\gamma^2 + 4d\gamma\lambda + 3\lambda^2)}\\
	&\equiv \frac{W_3(\alpha)}{k^6} + \frac{W_4(\alpha)}{k^2},\nonumber\\
	\langle |\hat{\theta}(\kv)|^2\rangle &= \frac{D_\theta}{d k^2}.\nonumber\\
\end{aligned}
\right.
\ea
where $k=\sqrt{k_x^2+k_y^2}$ and $\alpha$ is the polar angle such that $(k_x,k_y)=k(\cos\alpha,\sin\alpha) $. The constants are:
\ba
\label{W-1234-def}
\begin{aligned}
	W_1(\alpha) &= \frac{b^2 D_\theta (d\gamma + 4\lambda) \sin^2 2\alpha }{ 12 d \lambda (d^2 \gamma^2 + 4 d \gamma \lambda + 3\lambda^2)},\\
	W_2(\alpha) &= \frac{\gamma D (2 - \cos2\alpha) }{ 3\lambda },\\
	W_3(\alpha) &= \frac{b^2 D_\theta\Big[ d\gamma(15 + 8\cos2\alpha + \cos4\alpha) + \lambda(36 + 32\cos2\alpha + 4\cos4\alpha) \Big]}{ 24 d \lambda (d^2 \gamma^2 + 4 d \gamma \lambda + 3 \lambda^2)},\\
	W_4(\alpha) &= \frac{\gamma D (2 + \cos2\alpha) }{ 3 \lambda }.
\end{aligned}
\ea

\begin{figure}[h]
	\centering
	\includegraphics[width=5in]{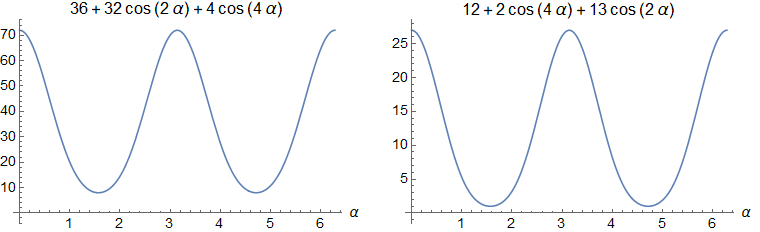}
	\caption{ The functions $W_3(\alpha)$ and $W_5(\alpha)$ are always positive in the interval $[0,2\pi]$.}
	\label{fig::W3W5}
\end{figure}

It is obvious that all the above coefficients $W_1(\alpha)$, $W_2(\alpha)$, $W_3(\alpha)$ (See Fig.~\ref{fig::W3W5}), $W_4(\alpha)$ are positive, except that $W_1(\alpha)$ might be zero at $\alpha=0,\pi/2, \pi, 3\pi/2$. 
We may conclude that as $k\rightarrow0$ in AD-I, $\langle |\hat{u}_x|^2\rangle$ scales as $k^{-6}$, $\langle |\hat{\theta}|^2\rangle$ scales as $k^{-2}$, and $\langle |\hat{u}_y|^2\rangle$ scales as $k^{-4}$. However, along the two special orthogonal axis $\hat{k}_x$ ($\alpha=0$ or $\pi$) and $\hat{k}_y$ ($\alpha=\pi/2$ or $3\pi/2$), $\langle |\hat{u}_x|^2\rangle$ scales as $k^{-2}$. To be more specific, along the two orthogonal axis $\hat{k}_x$ and $\hat{k}_y$ the correlation functions are,

\begin{alignat}{2}
	&\langle |\hat{u}_x(k_x,0)|^2\rangle = \frac{\gamma D}{3 \lambda k_x^2}, &\quad &\langle |\hat{u}_x(0,k_y)|^2\rangle = \frac{\gamma D}{\lambda k_y^2},\label{S_uy_model1_0}\\
	&\langle |\hat{u}_y(k_x,0)|^2\rangle = \frac{b^2 D_\theta}{d \lambda (d \gamma + \lambda) k_x^6} + \frac{\gamma D}{\lambda k_x^2}, &\quad &\langle |\hat{u}_y(0,k_y)|^2\rangle = \frac{b^2 D_\theta}{3 d \lambda (d \gamma + 3 \lambda) k_y^6} 
	+ \frac{\gamma D}{3\lambda k_y^2},\label{S_uy_model1_2}\\
	&\langle |\hat{\theta}(k_x,0)|^2\rangle = \frac{D_\theta}{d k_x^2}, &\quad &\langle |\hat{\theta}(0,k_y)|^2\rangle = \frac{D_\theta}{d k_y^2}\label{S_theta_ADI}.
\end{alignat}

In AD-II, after implementing Fourier Transform to the spatial component, we obtain
\ba
\left(\begin{matrix}
	\dot{\hat{u}}_x\\\dot{\hat{u}}_y\\\dot{\hat{\theta}}
\end{matrix}\right)+\mathbf{\Gamma}\left(\begin{matrix}
	\hat{u}_x\\\hat{u}_y\\\hat{\theta}
\end{matrix}\right)=\left(\begin{matrix}
	\sqrt{2D}\ \hat{\xi}_x\\\sqrt{2D}\ \hat{\xi}_y\\\sqrt{2D_\theta}\ \hat{\eta}
\end{matrix}\right).
\ea
where
\ba
\mathbf{\Gamma}=\frac{1}{\gamma}\left(\begin{matrix}
	\lambda(3k_x^2+k_y^2) & 2\lambda k_xk_y & 0\\2\lambda k_xk_y & \lambda(3k_y^2+k_x^2) & -b\\2\gamma\lambda c k_xk_y & \gamma\lambda c(3k_y^2+k_x^2) & 0
\end{matrix}\right).
\ea
The solution is
\begin{footnotesize} 
\ba
\left\{
\begin{aligned}
	\langle |\hat{u}_x(\kv)|^2\rangle &= \frac{b\left[16 D_\theta k_x^2 k_y^2 + c^2 D (k_x^4+12k_x^2k_y^2 + 27 k_y^4)\gamma^2\right] + 12c D k^4 (k_x^2 + 3k_y^2)\gamma\lambda}{3c k^4 \lambda\left[bc(k_x^2 + 9k_y^2)\gamma + 12k^4\lambda\right]}\\
	&\equiv \frac{W_5(\alpha)}{k^2} + \mathcal{O}(1),\nonumber\\
	\langle |\hat{u}_y(\kv)|^2\rangle &= \frac{3b^2 c D_\theta k^2 \gamma + b(3k_x^2 + k_y^2)\left[4D_\theta(3k_x^2 + k_y^2) + c^2 D (k_x^2 + 9k_y^2)\gamma^2\right]\lambda + 12 c D k^4 (3k_x^2+k_y^2) \gamma \lambda^2}{3 c k^4 \lambda^2\left[bc(k_x^2 + 9k_y^2)\gamma + 12 k^4 \lambda\right]}\\
	&\equiv \frac{W_6(\alpha)}{k^4} + \mathcal{O}(\frac{1}{k^2}),\nonumber\\
	\langle |\hat{\theta}(\kv)|^2\rangle &= \frac{b^2c^2D_\theta(k_x^2 + 3k_y^2)\gamma^2 + bck^2\gamma\left[13D_\theta k^2 + c^2 D (k_x^2 + 9k_y^2)\gamma^2\right]\lambda + 12k^6 (D_\theta + c^2 \gamma^2 D )\lambda^2}{ b c k^2 \lambda \left[bc(k_x^2 + 9k_y^2)\gamma + 12k^4 \lambda\right]}\\
	&\equiv \frac{W_7(\alpha)}{k^2} + \mathcal{O}(1).\nonumber\\
\end{aligned}
\right.
\ea
\end{footnotesize}
where
\ba
\label{W-567-def}
\begin{aligned}
	W_5(\alpha) &= \frac{ c^2\gamma^2 D(12 + 2\cos4\alpha + 13\cos2\alpha) +D_\theta(2-2\cos4\alpha) }{3 c^2 \gamma \lambda (5-4\cos2\alpha)},\\
	W_6(\alpha) &= \frac{b D_\theta}{c \lambda^2 (5-4\cos2\alpha)},\\
	W_7(\alpha) &= \frac{\gamma D_\theta (2-\cos2\alpha)}{\lambda(5-4\cos2\alpha)}.
\end{aligned}
\ea

It is obvious that all the coefficients $W_5(\alpha)$ (See Fig.~\ref{fig::W3W5}), $W_6(\alpha)$ and $W_7(\alpha)$ are always positive. We may conclude that as $k\rightarrow0$ in AD-II, $\langle |\hat{u}_x|^2\rangle$ and $\langle |\hat{\theta}|^2\rangle$ scale as $k^{-2}$, while $\langle |\hat{u}_y|^2\rangle$ scales as $k^{-4}$. Specially, along the two orthogonal axis $\hat{k}_x$ ($\alpha=0$ or $\pi$) and $\hat{k}_y$ ($\alpha=\pi/2$  or $3\pi/2$) the phonon correlation functions are,
\begin{alignat}{2}
	&\langle |\hat{u}_x(k_x,0)|^2\rangle = \frac{\gamma D}{3\lambda k_x^2}, &\quad &\langle |\hat{u}_x(0,k_y)|^2\rangle=\frac{\gamma D}{\lambda k_y^2},\label{S_uy_model2_0}\\
	&\langle |\hat{u}_y(k_x,0)|^2\rangle = \frac{bD_\theta}{c\lambda^2k_x^4}+\frac{\gamma D }{\lambda k_x^2}, &\quad &\langle |\hat{u}_y(0,k_y)|^2\rangle=\frac{bD_\theta}{9c\lambda^2k_y^4}+\frac{\gamma D}{3\lambda k_y^2},\label{S_uy_model2_2}\\
	&\langle |\hat{\theta}(k_x,0)|^2\rangle = \frac{\gamma D_\theta}{\lambda k_x^2} + \frac{D_\theta+c^2\gamma^2 D}{bc}, &\quad &\langle |\hat{\theta}(0,k_y)|^2\rangle=\frac{\gamma D_\theta}{3\lambda k_y^2} + \frac{D_\theta+c^2\gamma^2 D}{bc}\label{S_theta_ADII}.
\end{alignat}

In the limit of $b=0$, the phonon correlation functions in both AD-I and AD-II reduce to the same form, which is exactly the phonon correlation of passive crystals:
\ba
\langle |\hat{u}_x(\kv)|^2\rangle=\frac{\gamma D(2-\cos2\alpha)}{3\lambda k^2},\\
\langle |\hat{u}_y(\kv)|^2\rangle=\frac{\gamma D(2+\cos2\alpha)}{3\lambda k^2}.
\ea
and along the $\hat{k}_x$ and $\hat{k}_y$ axis, 
\ba
&&\langle |\hat{u}_x(k_x,0)|^2\rangle = \frac{\gamma D}{3 \lambda k_x^2},\quad \langle |\hat{u}_x(0,k_y)|^2\rangle = \frac{\gamma D}{\lambda k_y^2},\label{factor_ux_b0}\\
&&\langle |\hat{u}_y(k_x,0)|^2\rangle=\frac{\gamma D}{\lambda k_x^2},\quad \langle |\hat{u}_y(0,k_y)|^2\rangle= \frac{\gamma D}{3\lambda k_y^2}.\label{factor_uy_b0}
\ea
These results are consistent with the displacement correlation functions of isotropic solids~\cite{chaikin1995principles}:
\ba
\langle|\hat{u}_i(\mathbf{k})\hat{u}_j(-\mathbf{k})|\rangle=\frac{T\hat{k}_i\hat{k}_j}{3(\lambda+2\mu) k^2}+\frac{T(1-\hat{k}_i\hat{k}_j)}{\mu k^2},\label{Eqn::isotropicSolid}
\ea
by choosing $\lambda=\mu$ and $i=j$.

\section{Contour plots of the phonon correlation functions in the Collective Moving phase}
\begin{figure}[t!]
	\centering
	\includegraphics[width=6in]{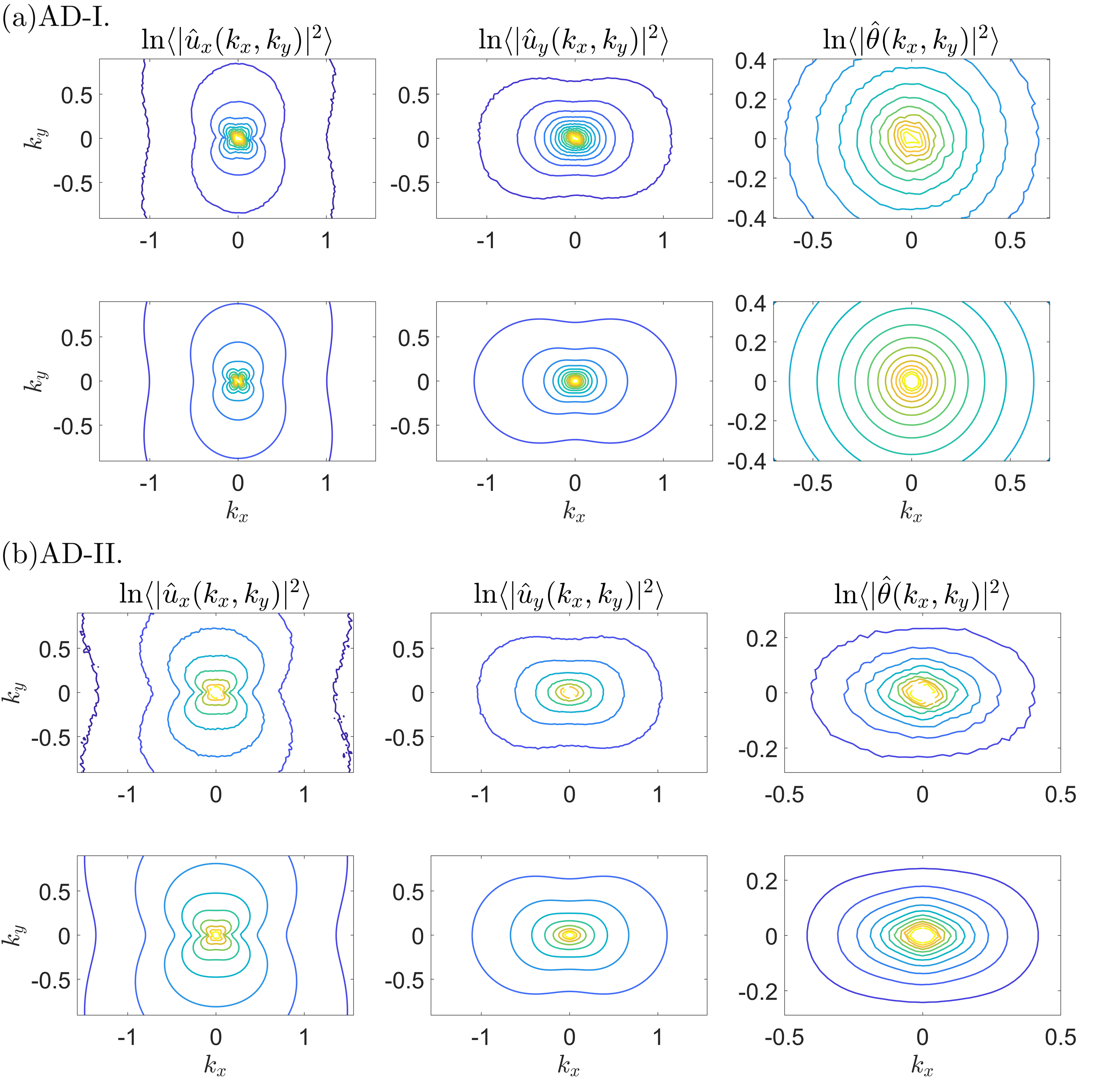}
	\caption{Contour plot of the logarithm of the phonon correlation functions $\langle|\hat{u}_x|^2\rangle$, $\langle|\hat{u}_y|^2\rangle$ and $\langle|\hat{\theta}|^2\rangle$ in $k$-space (arbitrary unit) in the Collective Moving phase, obtained from both simulation (first row in each subplot) and theoretical (second row in each subplot) results. (a) AD-I with $\tilde{d}=\tilde{d}_L=10$. (b) AD-II with $\tilde{c}=\tilde{c}_L=1$.}
	\label{k_space_profile}
\end{figure}

In Fig.~\ref{k_space_profile} we plot the contour curves of the phonon correlation functions in the collective moving phase in $k_x$-$k_y$ plane. The parameters are: $\tilde{d}_L=\tilde{d}=10$ and $\tilde{\kappa}=100$ for AD-I, $\tilde{c}_L=\tilde{c}=1$ and $\tilde{\kappa}=20$ for AD-II, $\tilde{b}_L=\tilde{b}=2$, $\tilde{D}_\theta^L=\tilde{D}_\theta=0.05$, $\tilde{\lambda}=\tilde{\mu}=\tilde{\kappa}\sqrt{3}/4$. Note that we let $\tilde{d}_L=\tilde{d}$ even though there is no direct correspondence between these two coefficients. The numerical results are given in the first row within each subplot of Fig.~\ref{k_space_profile} and the analytical results in the second row for comparison. The absolute value between the simulation and analytical curves might not be exactly equal, since they may differ by some constants such as the Fourier coefficients during FFT, which however are not of importance for our purpose. It can be seen that the scaling behaviors predicted by theory is consistent with the simulation results.
Besides, $\langle|\hat{\theta}|^2\rangle$ in AD-I is rotationally invariant, which originates from the rotational invariance of the active dynamics of $\theta$ in AD-I.  

\section{Generalization of the equations of motion into three dimensions}
In this section we discuss the analytical treatment of Eq.~(3a) in three dimensions, which can be written as
\ba
F_i=\mu \sum_{j=1}^{3} \partial_j^{2} u_{i}+\left(\lambda + \mu\right) \sum_{j=1}^{3} \partial_i\partial_j u_{j}
\ea
or more explicitly as
\ba
\left\{\begin{array}{ll}
	F_{x}&=  \mu(\partial_x^{2} u_{x} + \partial_y^{2} u_{x} + \partial_z^{2} u_{x}) + (\lambda+\mu)(\partial_x^{2} u_{x} + \partial_x\partial_y u_{y} + \partial_x\partial_z u_{z}), \\
	F_{y}&= \mu(\partial_x^{2} u_{y} + \partial_y^{2} u_{y} + \partial_z^{2} u_{y}) + (\lambda+\mu)(\partial_y\partial_x u_{x} + \partial_y^2 u_{y} + \partial_y\partial_z u_{z}), \\
	F_{z}&= \mu(\partial_x^{2} u_{z} + \partial_y^{2} u_{z} + \partial_z^{2} u_{z}) + (\lambda+\mu)(\partial_z\partial_x u_{x} + \partial_z\partial_y u_{y} + \partial_z^2 u_{z}).
\end{array}\label{eq:explicitElasticForce3d}
\right.
\ea
Under the assumption of $\theta\ll1$, we may decompose the active force direction into one parallel to the collective moving direction $\xh$ and the other perpendicular to it, i.e.,  $\hat{n}=\hat{n}_\parallel + \hat{n}_\perp$ where $|\hat{n}_\perp|\ll1$. We can also decompose the elastic force $\mathbf{F}=\mathbf{F}_\parallel+\mathbf{F}_\perp$. The overdamped dynamics of the displacement field is,
\ba
\gamma\dot{\mathbf{u}}(\mathbf{r},t)=b\ \hat{n}_\perp + \mathbf{F}(\mathbf{r},t)+\gamma\sqrt{2D}\hat{\xi}(\mathbf{r},t).\label{eq:overdamped3d}
\ea
where $\hat{\xi}(\mathbf{r},t)$ is a vector Gaussian white noise satisfying $\langle\xi_i(\mathbf{r},t)\rangle=0,\ \langle\xi_i(\mathbf{r},t)\xi_j(\mathbf{r}',t')\rangle=\delta_{ij}\delta(\mathbf{r}-\mathbf{r}')\delta(t-t')$, and it can be decomposed as $\hat{\xi}=\xi_x\xh + \hat{\xi}_\perp$. The continuum limit of (AD-I) and (AD-II) are respectively,
\ba
\begin{aligned}
	\dot{\hat{n}}_\perp(\mathbf{r},t) &= d\ \Delta \hat{n}_\perp + \sqrt{2D_\theta}\ \hat{\eta}_\perp(\mathbf{r},t),\qquad \qquad \nonumber
	\text{(AD-I')}\\
	\dot{\hat{n}}_\perp(\mathbf{r},t) &= c\ (\hat{n}\times\mathbf{F})\times\hat{n} + \sqrt{2D_\theta}\ \hat{\eta}_\perp(\mathbf{r},t) \\
	&\approx c\ \mathbf{F}_\perp + \sqrt{2D_\theta}\ \hat{\eta}_\perp(\mathbf{r},t).\quad\qquad\qquad \text{(AD-II')}
	\nonumber
\end{aligned}
\ea
where $\hat{\eta}_\perp(\mathbf{r},t)$ is a 2d vector Gaussian white noise in the $\yh$-$\zh$ plane satisfying $\langle\eta_{\perp i}(\mathbf{r},t)\rangle=0,\ \langle\eta_{\perp i}(\mathbf{r},t)\eta_{\perp j}(\mathbf{r}',t')\rangle = \delta_{ij}\delta(\mathbf{r}-\mathbf{r}')\delta(t-t')$.

We can substitute Eq.($\ref{eq:explicitElasticForce3d}$) into Eq.($\ref{eq:overdamped3d}$) together with Eq.(AD-I'),  and obtain the dynamic equations for AD-I:
\ba
\left\{\begin{array}{ll}
	\gamma \, \dot{u}_x&=F_x+\gamma\sqrt{2D} \, \xi_x\\
	\gamma \, \dot{\mathbf{u}}_\perp&=b\ \hat{n}_\perp + \mathbf{F}_\perp + \gamma\sqrt{2D}\, \hat{\xi}_\perp\\
	\dot{\hat{n}}_\perp &= d\ \Delta \hat{n}_\perp + \sqrt{2D_\theta}\ \hat{\eta}_\perp
	\label{eq:peturb_model1_3d}
\end{array}.
\right.
\ea

For AD-II, the dynamic equations are given by,
\ba
\left\{\begin{array}{ll}
	\gamma \, \dot{u}_x&=F_x+\gamma\sqrt{2D} \, \xi_x\\
	\gamma \, \dot{\mathbf{u}}_\perp&=b\ \hat{n}_\perp + \mathbf{F}_\perp + \gamma\sqrt{2D}\, \hat{\xi}_\perp\\
	\dot{\hat{n}}_\perp&= c\ \mathbf{F}_\perp + \sqrt{2D_\theta}\ \hat{\eta}_\perp
	\label{eq:peturb_model2_3d}
\end{array}.
\right.
\ea

It can be easily seen that Eqs.~(\ref{eq:peturb_model1_3d}) and (\ref{eq:peturb_model2_3d})
are direct generalizations of Eqs.~(5) and (8) to the 3d case. The above arguments can be easily extended to higher dimensions.

\end{widetext}
